\documentclass[prb,aps,twocolumn,amsmath,amssymb,floatfix,superscriptaddress]{revtex4}

\usepackage{color}
\usepackage{soul}
\usepackage[colorlinks=true, citecolor=blue, urlcolor=blue]{hyperref}
\usepackage{graphics,epstopdf}
\usepackage{epsf,graphics,graphicx}

\begin{document}

\title{Spin-dependent electron transfer through a ring-wire coupled junction: Role of in-plane electric field}

\author{Prabhab Patra}

\affiliation{Physics and Applied Mathematics Unit, Indian Statistical Institute, 203 Barrackpore Trunk Road, Kolkata-700 108, India}

\author{Santanu K. Maiti}

\email{santanu.maiti@isical.ac.in}

\affiliation{Physics and Applied Mathematics Unit, Indian Statistical Institute, 203 Barrackpore Trunk Road, Kolkata-700 108, India}

\begin{abstract}

We study spin-dependent transport in a hybrid magnetic system, where a non-magnetic (NM) wire is coupled to a side-attached antiferromagnetic (AFM) mesoscopic ring, placed between two non-magnetic electrodes subject to an in-plane electric field oriented perpendicular to the NM wire. The system is described within a tight-binding (TB) framework, and transport properties are computed using the non-equilibrium Green's function (NEGF) formalism.
We consider two junction configurations distinguished by the ring-wire coupling: a single-coupled junction and a double-coupled junction. In the single-coupled configuration, the coupling geometry alone breaks the spin symmetry, yielding a finite spin polarization (SP) even without any external field. The in-plane electric field further enhances the symmetry breaking in both configurations, serving as an efficient tuning parameter that drives the SP nearly $100\%$ in the low-bias region. In the double-coupled configuration, spin symmetry is preserved in the absence of the external field, and the electric field acts as a sole source of symmetry breaking, producing a large SP.  
Finite temperature effects and different system sizes are examined, confirming the robustness of the observed features. To validate the findings over a wide parameter space, we considered different sets of parameters and found that the key signatures remain unchanged. Our results demonstrate that such hybrid structures are promising candidates for realizing an externally controllable spintronic device in low-dimensional systems.

\end{abstract}

\maketitle

\section{Introduction}

Spin-dependent transport in low-dimensional systems has attracted significant attention due to its relevance for spintronic applications\cite{stronics1,stronics2,stronics3}. The discovery of giant magneto-resistance (GMR) opened the door to spintronics, which 
quickly gained recognition as a compelling area of research within mesoscopic physics. Spintronics exploits the electron spin degree of 
freedom to process and store information, offering the possibility of reduced power consumption and enhanced device functionality\cite{d1,d2,d3,d4,d5,d6}. One of the most important concerns in spintronics research is finding dependable ways to generate 
and control spin-polarized currents effectively. The central bridging element of a simple device plays a crucial role in spin splitting, 
so it is very important to choose an efficient spin-dependent functional element.

In the early days, the most common approach was to use a ferromagnetic\cite{fm1,fm2,fm3,afm1} material as the central element, due to 
its non-zero net magnetization, which acts as a natural spin filter. However, ferromagnetic materials as bridging elements also bring some serious limitations. The electron injection from the electrode to the functional element is disrupted because a large resistance\cite{r1,r2} appears at the junction interfaces. Moreover, ferromagnetic materials contain stray fields, which affect neighboring devices and slow down the switching speed.
To resolve these issues, researchers began exploring alternative materials. One important direction is the spin-orbit coupled (SOC) system\cite{so1,so2,so3,so4,so5,so6,so7}, which can be controlled electrically without any external magnetic field. Primarily, two types 
of SOCs are relevant: Rashba\cite{so1} and Dresselhaus\cite{so2} SOC. The origin of the Rashba SOC is structural asymmetry, whereas the
Dresselhaus SOC arises from bulk crystal asymmetry. Rashba SOC is generally preferred due to its external tunability. However, SOC systems
cannot achieve large spin polarization due to their weak coupling\cite{ht} strength.

Antiferromagnetic (AFM) materials can help circumvent all these limitations as spin-dependent functional elements. AFM-based devices can 
operate up to very high frequency limits. In recent years, AFM spintronics has emerged as an active and rapidly growing research
direction\cite{af2,af3,af4,af5,af6,af7,af8}. Unlike ferromagnets, AFM materials are stable even when unwanted magnetic forces act on them 
from outside and produce no stray fields, making them highly suitable for dense device integration\cite{af3,af8}. Several theoretical and
experimental works have demonstrated spin-dependent transport, spin Hall effects, and spin-transfer torque phenomena in AFM-based
systems\cite{af9,st1,st2}. Furthermore, the electrical detection and manipulation of AFM order has been demonstrated experimentally\cite{st3},
opening the possibility of purely electrically controlled AFM spintronic devices. Despite these promising features, \textit{exploiting AFM
materials as active spin-filtering elements in mesoscopic hybrid junctions remains relatively unexplored, particularly in the context of
low-dimensional tight-binding models where the role of geometry and coupling can be studied systematically}.

Among the various hybrid geometries proposed for spin filtering, the combination of a non-magnetic conducting wire coupled to a side-attached
mesoscopic ring has received considerable attention. When the ring is antiferromagnetic, such hybrid junctions have been studied in the 
context of equilibrium circulating currents\cite{wr1,wr2,wr23}. However, spin-polarized transport through such NM wire-AFM ring junctions 
has never been addressed. This is a non-trivial problem: unlike ferromagnetic rings, an AFM ring carries zero net magnetization, and it is 
far from obvious how spin polarization can emerge in such a system. The only comparable study of spin transport in ring-wire junctions under 
an in-plane electric field considered a ferromagnetic ring\cite{wr3}, where SP is an expected outcome due to net magnetization. Extending 
this idea to an AFM ring raises a fundamental question: can spin polarization be achieved in a clean system with zero net magnetization, 
and if so, what controls it? In the present work, we address this question. We show that an in-plane electric field acting on the AFM ring
serves as an efficient external tuning parameter for SP. In addition to the role of the electric field, we investigate two different junction
configurations distinguished by the number of ring-wire couplings connecting the AFM ring to the NM wire. More remarkably, we demonstrate 
that even without any external field, the geometry of the ring-wire coupling alone determines whether spin symmetry is broken or preserved.

Specifically, we consider an NM wire coupled to a side-attached AFM mesoscopic ring, placed between two non-magnetic electrodes. The 
magnetic moments of the AFM ring are aligned alternately along the quantized $+Z$ and $-Z$ directions. Consequently, the total Hamiltonian 
$H$ (say) can be decoupled into two sub-Hamiltonians, namely $H=H_{\uparrow}+H_{\downarrow}$. To break the spin symmetry, an {\em in-plane} 
external electric field is applied perpendicular to the NM wire. This external field affects only the AFM ring and has no direct influence 
on the NM chain. The role of the in-plane field is crucial, as it not only provides external control over the spin polarization but is also
solely responsible for breaking the spin symmetry in the double-site coupling configuration. Importantly, the field introduces 
a deterministic and smooth modulation of the ring site energies, whose influence on spin polarization (SP)\cite{sp1,sp2} is qualitatively
different from that of conventional disorder which is commonly used in usual statistical sense. 

A tight-binding\cite{TB1,TB2,TB3,TB4,TB5} framework is used to model the electrode--ring-wire--electrode nano-junction. The spin-resolved
(up, down) transmission spectra are computed within the NEGF formalism\cite{negf1,negf2,negf3,negf4,negf5,sy1}, and the corresponding
spin-dependent currents are calculated using the Landauer-B\"{u}ttiker prescription\cite{LB1}. The roles of the ring-wire coupling and 
the in-plane electric field on SP are analyzed systematically. We analyze all the results for two different junction configurations depending 
on the ring-wire coupling geometry. The obtained results provide useful insights into spin-dependent transport and its external control in
AFM-based hybrid nanostructures. The effect of finite temperature is also discussed to assess the thermal stability of the observed features.
Furthermore, the consistency of our results is verified by considering different system sizes, Fermi energies, in-plane electric field
strengths, ring-wire coupling strengths, and bias voltages.

The remaining part of the work is arranged as follows. Section II describes the model and the theoretical background we used for studying 
spin-dependent transport phenomena. Section III presents all the numerical results and discussion, while Section IV provides an experimental
perspective of our proposed model nanojunction. Section V contains the summary of our findings. 

\section{MODEL AND THEORETICAL FRAMEWORK}

This section describes our model nanojunction and all the required tight-binding Hamiltonians. It also illustrates all the necessary 
formulae for the studied quantities in detail.

\subsection{Junction setup and the Hamiltonian}

We first describe the proposed hybrid junction, which consists of a clean NM wire and a side-attached impurity-free 
\begin{figure}[htbp]
\centering
\resizebox{8.5cm}{4cm}{\includegraphics{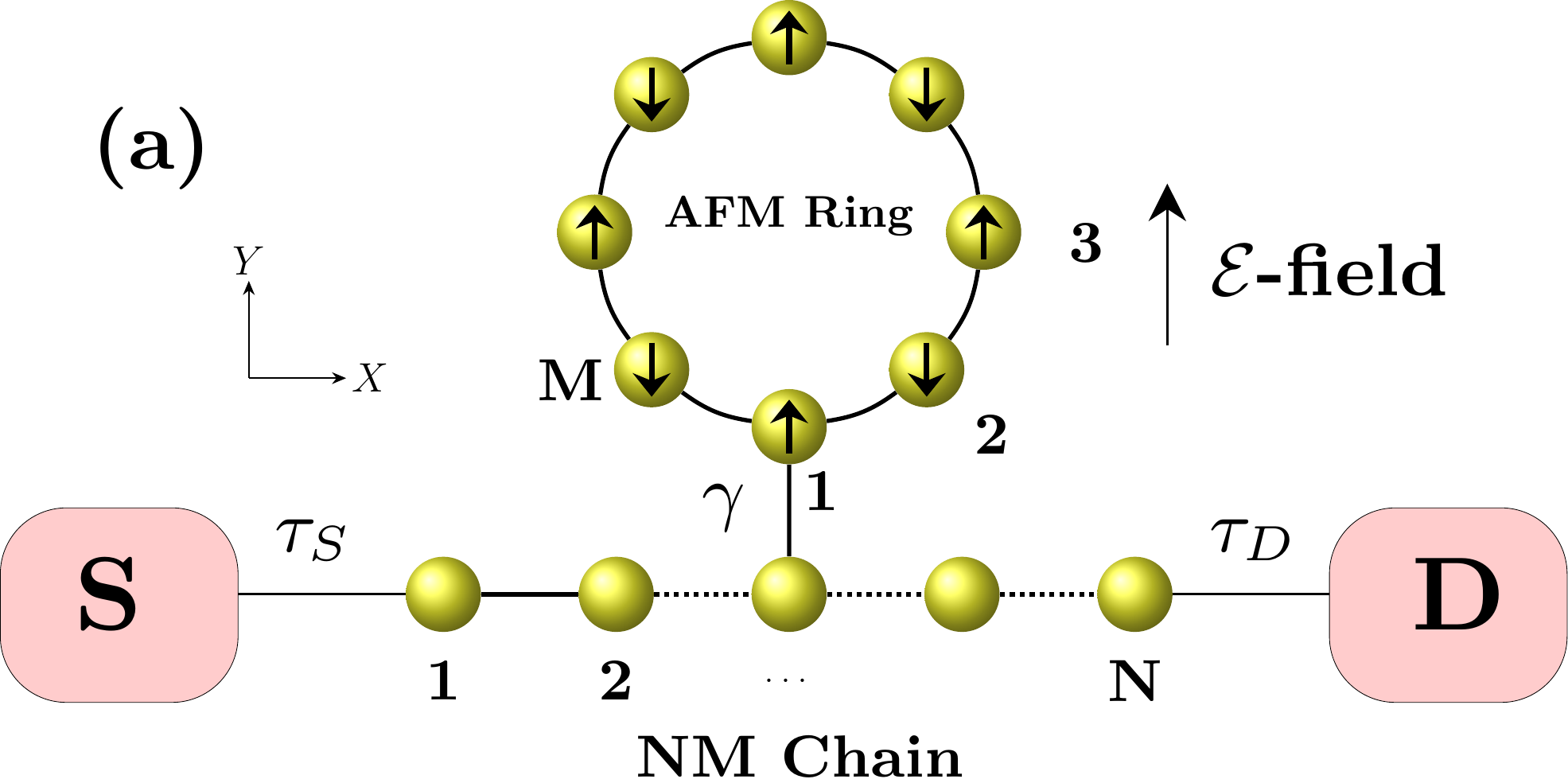}}\\[1cm]
\resizebox{8.5cm}{4cm}{\includegraphics{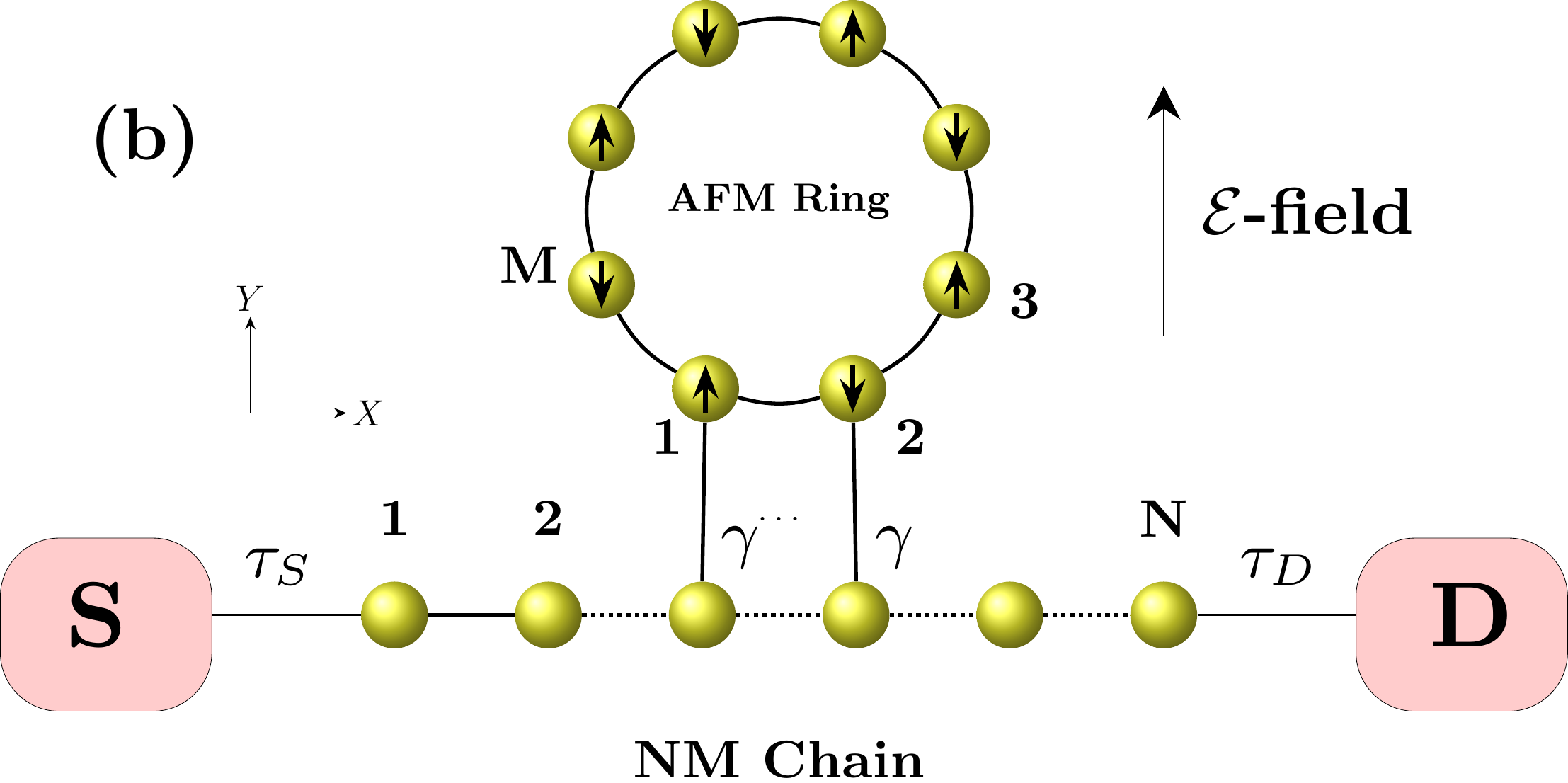}}
\caption{(Color online). A schematic representation of the junction setup. A non-magnetic wire is connected to a side-attached 
antiferromagnetic ring, positioned between the source (S) and drain (D) electrodes. Depending upon the number of ring-wire couplings, 
two distinct junction configurations are considered: (a) single-coupling (junction 1) and (b) double-coupling (junction 2). Along with 
this, an external in-plane electric field of strength $\mathcal{E}$, acting perpendicular to the NM wire, is applied to the system.}
\label{fig:f1}
\end{figure}
AFM ring symmetrically coupled to the NM wire through a single coupling. The local magnetic moments of the AFM ring are aligned alternately
along the $+Z$ and $-Z$ directions, which define the spin quantization axis. Two ideal, semi-infinite, one-dimensional electrodes with no
backscattering, namely the source (S) and drain (D), are connected to the two ends of the NM wire. An in-plane external electric field is
applied to the central hybrid region perpendicular to the NM wire as shown in Fig.~\ref{fig:f1}(a). Similarly, another junction setup is
considered, which is identical to the previous one, except that here two adjacent ring-wire couplings between the ring and the chain are 
taken as shown in Fig.~\ref{fig:f1}(b).

The complete Hamiltonian describing the junction is written as
\begin{equation}
\mathcal{H}=H_{rw}+H_S+H_D+H_{cpl}
\label{eq1}
\end{equation}
where the first sub-Hamiltonian $H_{rw}$ is associated with the ring-wire coupled system, which is decomposed as $H_{rw}=H_r+H_w+H_{cl}$. 
$H_r$ and $H_w$ describe the Hamiltonians of the AFM ring and the NM wire, respectively. The coupling between ring and wire is incorporated 
by $H_{cl}$. The second and third sub-Hamiltonians $(H_S, H_D)$ of Eq.~\ref{eq1} correspond to the source and drain. The coupling between
electrodes and NM wire is given by $H_{cpl}$. The explicit forms of the sub-Hamiltonians are provided below in the TB framework. The central
sub-Hamiltonian $H_{rw}$ for junction 1 is expressed as,
\begin{equation}
\begin{split}
    H_{rw}&= \sum_n \boldsymbol{c_n^{r\dagger} }(\boldsymbol{\epsilon_n^r}-\boldsymbol{\vec{h}_n.\vec{\sigma}})\boldsymbol{c_n^r}
    +\sum_n (\boldsymbol{c_{n+1}^{r\dagger}t_r c_n^r}+h.c.)\\
    &+\sum_n\boldsymbol{c_n^{w\dagger}\epsilon_n^wc_n^w}
    +\sum_n(\boldsymbol{c_{n+1}^{w\dagger}t_wc_n^w}+h.c.)\\
    &+\underbrace{(\boldsymbol{c_1^{r\dagger}\gamma c_p^w}+h.c.)}_{H_{cl}^{(1)}}.
    \raisetag{2.2\baselineskip}
\end{split}
\label{eq2}
\end{equation}
The first two terms of $H_{rw}$ characterize the AFM ring. Here $\boldsymbol{c_n^{r\dagger}}=\begin{pmatrix} c_{n\uparrow}^{r\dagger} & c_{n\downarrow}^{r\dagger} \end{pmatrix}$, $c_{n\sigma}^{r\dagger}$ and $c_{n\sigma}^{r}$ denote the operators responsible for the creation 
and destruction of an electron with spin $\sigma$ at site $n$, where $\sigma=\uparrow, \downarrow$. Similarly in the next term,
$c_{n\sigma}^{w\dagger}$ and $c_{n\sigma}^{w}$ are the creation and annihilation operators for the NM wire. Whereas $\boldsymbol{t_r}$ 
is a diagonal $(2\times2)$ matrix with diagonal elements $t_{r\uparrow}=t_{r\downarrow}=t_r$. $t_r$ denotes the nearest-neighbor hopping
(NNH) strength in the AFM ring. The effective site energy of the AFM ring is given by 
$(\boldsymbol{\epsilon_n^r}-\boldsymbol{\vec{h}_n \cdot \vec{\sigma}})$.
The spin-dependent scattering is described by the term\cite{scat} $\boldsymbol{\vec{h}_n \cdot \vec{\sigma}}$, where
$\boldsymbol{\vec{h}_n}=J\langle \vec{S}_n \rangle$. $J$ is the coupling strength and $\langle \vec{S}_n \rangle$ denotes the net 
local spin at site $n$. Here $\boldsymbol{\vec{\sigma}=(\sigma_x,\sigma_y,\sigma_z)}$ represents the Pauli matrices. The 
$\boldsymbol{\vec{h}_n \cdot \vec{\sigma}}$ accounts for the interaction between an itinerant electron and a local magnetic moment. 
In spherical polar coordinates, it can be expressed as, $\boldsymbol{\vec{h}_n \cdot \vec{\sigma}} = h_n \left( \sin\theta_n \cos\phi_n \, \boldsymbol{\sigma_x} + \sin\theta_n \sin\phi_n \, \boldsymbol{\sigma_y} + \cos\theta_n \, \boldsymbol{\sigma_z} \right)$. Here, 
$\theta_n$ and $\phi_n$ are the polar and azimuthal angles, respectively, defining the orientation of the local magnetic moment at site 
$n$. In our system, the magnetic moments of the ring are directed alternately in $+z$ and $-z$ directions, so 
$\boldsymbol{ \vec{h}_n \cdot \vec{\sigma} }=\text{diag}(h_n, -h_n)$ and site energy matrix of the AFM-ring is expressed as
$\boldsymbol{\epsilon_n^r}=\text{diag}(\epsilon_n^r,\epsilon_n^r)$, which is spin-independent. The site energy can be expressed as 
the combination of two contributions, $\epsilon_n^r=\epsilon_n^{r(\mathcal{E}=0)}+\epsilon_n^{r(\mathcal{E} \neq0)}$.
$\epsilon_n^{r(\mathcal{E}=0)}$ is the site energy of the ring in the absence of an external in-plane electric field. The second contribution
$\epsilon_n^{r(\mathcal{E} \neq0)}$ takes into account the effect of a finite in-plane electric field acting on the ring (perpendicular
to the NM wire). The explicit form is given by 
\begin{equation}
\epsilon_n^{r(\mathcal{E} \neq0)}=\left(\frac{eaN_r\mathcal{E}}{2\pi}\right)\cos\left(\frac{2\pi(n-1)}{N_r}\right)
\label{fldeq1}
\end{equation}
where $\mathcal{E}$ represents the electric field strength, and $e$, $N_r$, and $a$ refer to the electron charge, 
ring size, and the lattice 
spacing in the ring, respectively. The in-plane electric field results in a deterministic, site-dependent modulation 
of the ring site energies. For convenience, to make the in-plane electric field dimensionless, Eq.~\ref{fldeq1} can be rewritten as 
\begin{equation} 
\epsilon_n^{r(\varepsilon \neq0)}=\left(\frac{N_r \varepsilon t_r}{2\pi}\right)\cos\left(\frac{2\pi(n-1)}{N_r}\right)
\label{fldeq2}
\end{equation}
where, $\varepsilon$ ($=ea\mathcal{E}/t_r$) corresponds to the dimensionless field strength.

The 3rd and 4th terms of the ring-wire coupled Hamiltonian (Eq.~\ref{eq2}) describe the Hamiltonian of the NM wire. Its site-energy 
matrix takes the form, $\boldsymbol{\epsilon_n^w}=\text{diag}(\epsilon_n^w, \epsilon_n^w)$. Note that the site energy $\epsilon_n^w$ 
is spin-degenerate and remains unaffected by the in-plane electric field, since the field direction is perpendicular to the wire.
$\boldsymbol{t_w}=\text{diag}(t_w, t_w)$ is the nearest-neighbor hopping matrix of the NM wire, where $t_w$ is the corresponding 
hopping strength.

\vskip 0.2cm
\noindent 
{\em Explanation of site energy dependence of the ring and wire on electric field}: The effect of the electric field on the 
site energies can be clearly understood by considering the direction of the electric field with respect to the ring-wire junction setup. 
Our chosen system is placed in the $X$-$Y$ plane (see Fig.~\ref{fig:f1}). The wire is placed along the $X$ direction, and the $y$-coordinates 
of all its sites are assumed to be zero. The ring is coupled to the wire via a single or double connecting link, as described in the figure. 
For the sake of simplicity, we focus on junction 1. The same argument is also applicable to the junction 2 setup. The site $1$ of the ring 
is coupled to the NM wire. The electric field is applied perpendicular to the wire and is directed along the $Y$ axis. The setup can be 
realized in a laboratory by placing the model nanojunction between two parallel plates that produce a uniform electric field. Since the wire 
is placed at $y=0$, the site energies of the wire are not affected by the in-plane electric field. For a non-zero value of $y$, the site
energies of the wire are modified by the same amount, and thus, the physics remains unchanged. In contrast, depending on the orientation 
of different lattice sites in the ring, their site energies are modified differently by the electric field. The argument within the cosine
factor arises from the angular separation between the neighboring lattice sites measured from the downward vertical direction starting
from site number $1$ of the ring. In our calculation, we consider the center of the ring as the origin ($0, 0$).  
Equation~\ref{fldeq2} clearly shows that the site energies are deterministically modulated in the presence of the in-plane electric field.

The last term of Eq.~\ref{eq2} represents the coupling between the AFM ring and the NM wire, described by $H_{cl}^{(1)}$. In junction 1, 
the site $1$ of the ring is coupled to the $p$th site of the wire with coupling strength $\gamma$.

The source and drain electrodes are modeled through the Hamiltonians
\begin{equation}
    H_{S(D)}=\sum_n \boldsymbol{d_n^\dagger\epsilon_0d_n}
    +\sum_n(\boldsymbol{d_{n+1}^\dagger t_0d_n}+h.c.),
\end{equation}
where $\boldsymbol{\epsilon_0}=\text{diag}(\epsilon_0,\epsilon_0)$ and $\boldsymbol{t_0}=\text{diag}(t_0,t_0)$ represent the spin-independent
site-energy matrix and nearest-neighbor hopping matrix of the non-magnetic electrodes, with $\epsilon_0$ and $t_0$ being the corresponding
site energy and hopping strength, in that order. $\boldsymbol{d_n^{\dagger}}=\begin{pmatrix} d_{n\uparrow}^{\dagger} & d_{n\downarrow}^{\dagger} \end{pmatrix}$, $d_{n\sigma}^\dagger$ $(d_{n\sigma})$ creates (annihilates) an electron with spin $\sigma$ in the source and drain electrodes.

The coupling between the central element and the two electrodes takes the form of the sub-Hamiltonian $H_{cpl}$,
\begin{equation}
H_{cpl} = \boldsymbol{c_1^{w\dagger} \tau_S d_0} + 
\boldsymbol{c_{N_w}^{w\dagger} \tau_D d_{N_w+1}} + \text{h.c.}
\end{equation}
The electrodes are semi-infinite, with atomic sites running from $-\infty$ to $0$ for the source and from $N_w+1$ to $\infty$ for the drain. These electrodes (source and drain) are coupled to the NM wire at site 1 and $N_w$ with coupling strengths $\tau_S$ and $\tau_D$, respectively

For junction 2, all sub-Hamiltonians remain identical to those of junction 1. The only modification is in the coupling term $H_{cl}$, which 
for junction 1 is given by the last term of Eq. \ref{eq2}, and for junction 2 if takes the form:
\begin{equation}
H_{cl}^{(2)}= \boldsymbol{c_1^{r\dagger}\gamma c_p^w + c_2^{r\dagger}\gamma c_q^w} + \text{h.c.}
\end{equation}
Therefore, $H_{rw}$ for junction 2 is obtained by replacing $H_{cl}^{(1)}$ in Eq.~(2) with $H_{cl}^{(2)}$ above. Here, the site $1$ and $2$
of the AFM ring are coupled to two consecutive nearest-neighbor sites, $p$ and $q$, of the NM wire with equal coupling strength $\gamma$.

\subsection{Transmission probability: NEGF formalism}

In mesoscopic systems, the current is calculated with the help of the transmission probability. There are several techniques to determine the transmission probability; the non-equilibrium Green's function (NEGF) formalism\cite{negf1,negf2,negf3,negf4,negf5,sy1} is one of the more widely used approaches. Under this formalism, the probability that an electron leaving the source electrode with spin $\sigma$ and successfully reaching the drain electrode with spin $\sigma^{\prime}$ through the central region is figured out using the electrode self-energies $(\Sigma_S,\Sigma_D)$ together with the retarded and advanced Green's functions $(G^r,G^a)$.
The spin-resolved transmission then takes the form
\begin{equation}
T_{\sigma\sigma'}=\text{Tr}\left[\boldsymbol{\Gamma_S^\sigma G^r \Gamma_D^{\sigma'} G^a}\right].
\end{equation}
The self-energy matrices $\boldsymbol{\Sigma_S}$ and $\boldsymbol{\Sigma_D}$ effectively incorporate the influence of the semi-infinite electrodes on the central region. Since the electrodes are non-magnetic, the self energies are spin-independent and can be written as $\boldsymbol{\Sigma_{S(D)}}=\begin{pmatrix} \Sigma_{S(D)} &0\\ 0& \Sigma_{S(D)} \end{pmatrix}$, where the non-zero elements are given by
\begin{equation}
    \Sigma_{S(D)}= \frac{\tau_{S(D)}^2}{2t_0^2} \left[ (E-\epsilon_0) -i\sqrt{4t_0^2-(E-\epsilon_0)^2} \right],
\end{equation}
where $\Gamma_{S(D)}$ denotes the coupling matrix associated with the source (drain) electrode, and is directly connected to the imaginary part of the self-energy $\Sigma_{S(D)}$. The retarded Green's function is expressed as
\begin{equation}
\boldsymbol{G^r}=(E\boldsymbol{I}-H_{rw}-\boldsymbol{\Sigma_S}-\boldsymbol{\Sigma_D})^{-1}.
\end{equation}
Here, $\boldsymbol{I}$ is the unit matrix of the same order as the Hamiltonian of the central region. The advanced Green's function $\boldsymbol{G^a}$ can be obtained by taking the Hermitian conjugate of $\boldsymbol{G^r}$.

As the magnetic moments are aligned alternately along the $+Z$ and $-Z$ directions, the spin-flip transmission components vanish. 
From the above, the transmission probability corresponding to spin $\sigma$ is written as
\begin{equation}
T_{\sigma}=\text{Tr}\left[\boldsymbol{\Gamma_S^\sigma G^r \Gamma_D^{\sigma} G^a}\right].
\end{equation}
For the situation where the magnetic moments are not oriented along the spin quantization axis, but rather along a particular direction 
while maintaining zero net magnetization, the full Hamiltonian of the ring-wire system cannot be decoupled into two independent spin 
subspaces. Here, spin-flip transmission occurs together with spin-conserving transmission, and the net spin-up and spin-down transmission
probabilities are obtained from their respective components.

\subsection{Junction current}

After obtaining the transmission probability, the next step is to look at the spin-dependent current flowing through the junction. 
This is done through the Landauer-B\"uttiker prescription\cite{LB1}, which gives: 
\begin{equation}
I_{\sigma}(V)=\frac{e}{2\pi\hbar}\int T_{\sigma}(E)\left[f_S(E)-f_D(E)\right]dE.
\end{equation}
The factor $\hbar$ corresponds to the reduced Planck constant. The functions $f_S(E)$ and $f_D(E)$ represent the Fermi--Dirac distribution functions of the source and drain electrodes, and are expressed as
\begin{equation}
f_{S(D)}(E)=\frac{1}{1+\exp[(E-\mu_{S(D)})/k_B\Theta]}.
\end{equation}
Here, $\mu_{S(D)}$ stands for the electro-chemical potential of the source (drain), $k_B$ is the Boltzmann constant, and $\Theta$ refers to the absolute temperature. The electro-chemical potentials of the source and drain electrodes are expressed as
\begin{equation}
\mu_S = E_F + \frac{eV}{2} ~~ \mbox{and} ~~ \mu_D = E_F - \frac{eV}{2}
\end{equation}
where, $E_F$ is the equilibrium Fermi energy and $V$ is the bias voltage applied across the junction.

\subsection{Spin Polarization}

Using the spin-dependent currents, the degree of spin polarization (SP)\cite{SP1} can be evaluated as
\begin{equation}
SP=\frac{I_\uparrow-I_\downarrow}{I_\uparrow+I_\downarrow} \times 100\%.
\end{equation}
Here, $I_{\uparrow(\downarrow)}$ denotes the spin-up (spin-down) current through the system. A positive (negative) value of SP indicates 
the dominance of spin-up (spin-down) current through the junction. In the calculations presented here, the total current is always finite
within the bias window considered. Complete ($100\%$) spin polarization is achieved when one of the spin-current components vanishes.

\begin{figure*}[htbp]
\centering

\begin{minipage}{0.32\textwidth}
\centering
\includegraphics[width=\linewidth,height=4.2cm]{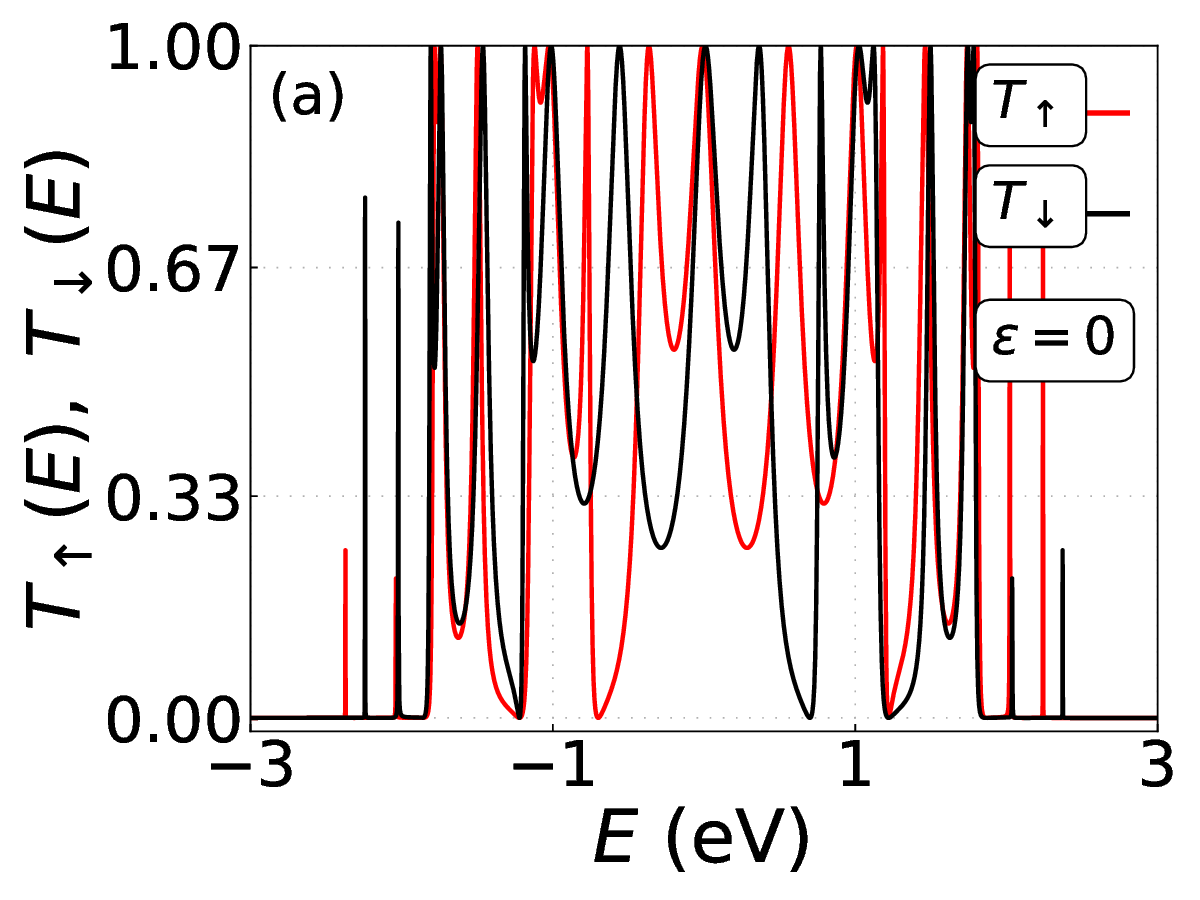}
\end{minipage}
\hfill
\begin{minipage}{0.32\textwidth}
\centering
\includegraphics[width=\linewidth,height=4.2cm]{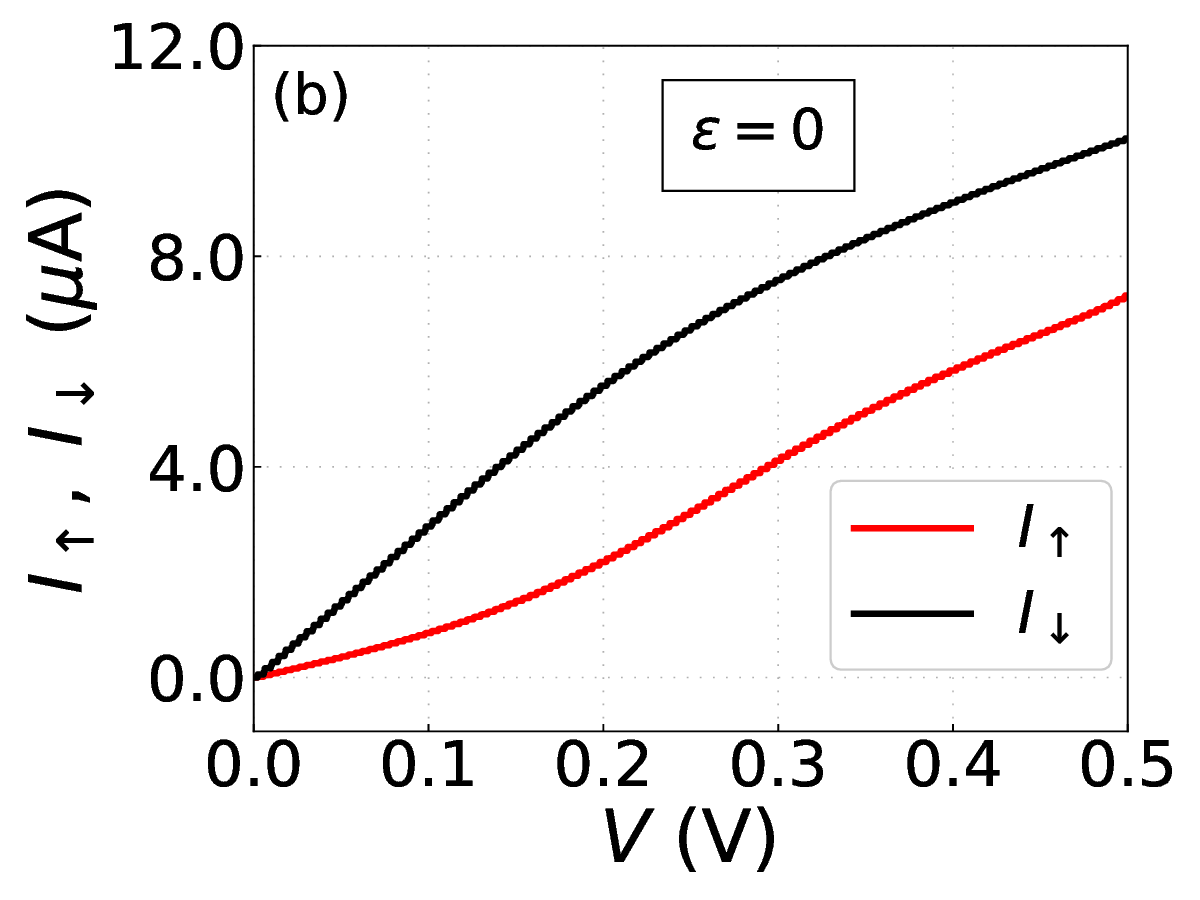}
\end{minipage}
\hfill
\begin{minipage}{0.32\textwidth}
\centering
\includegraphics[width=\linewidth,height=4.2cm]{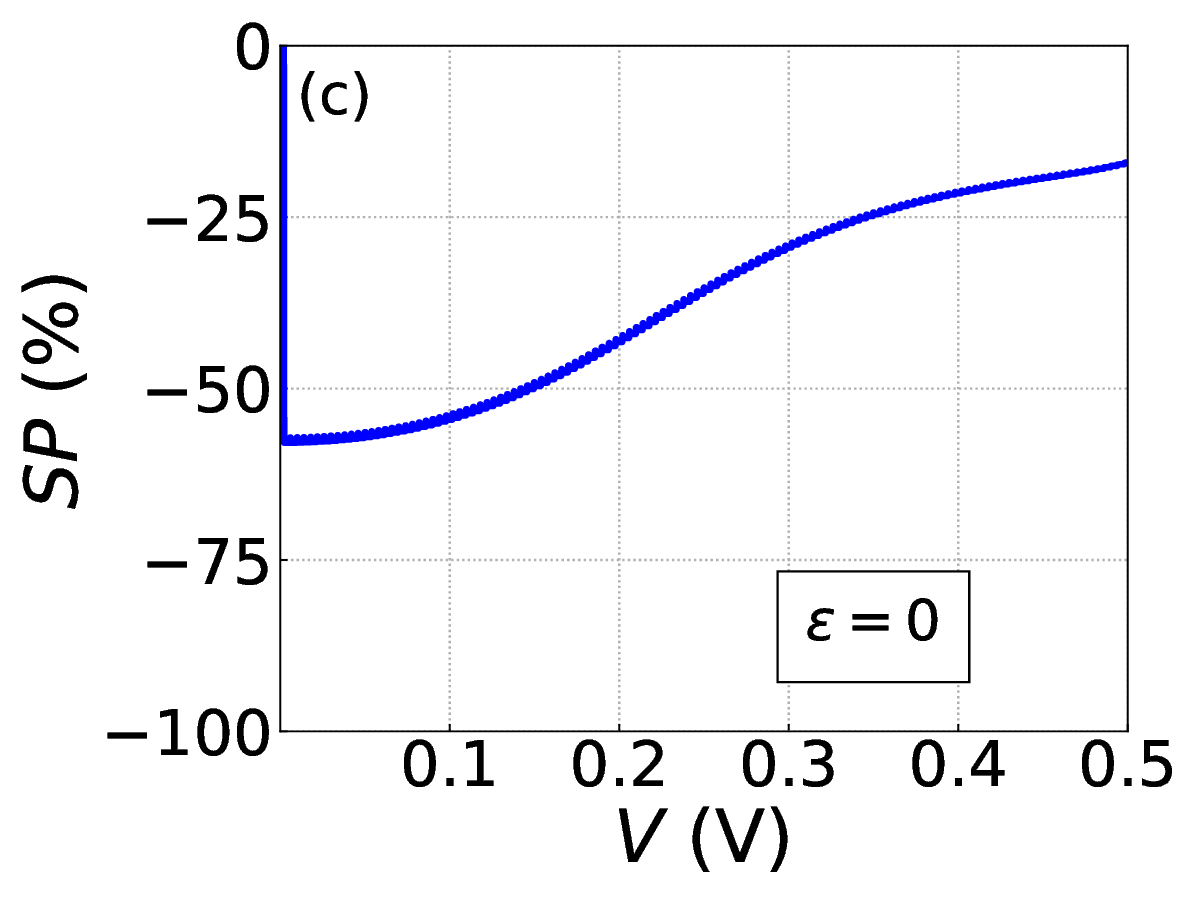}
\end{minipage}

\vspace{0.2cm}

\begin{minipage}{0.32\textwidth}
\centering
\includegraphics[width=\linewidth,height=4.2cm]{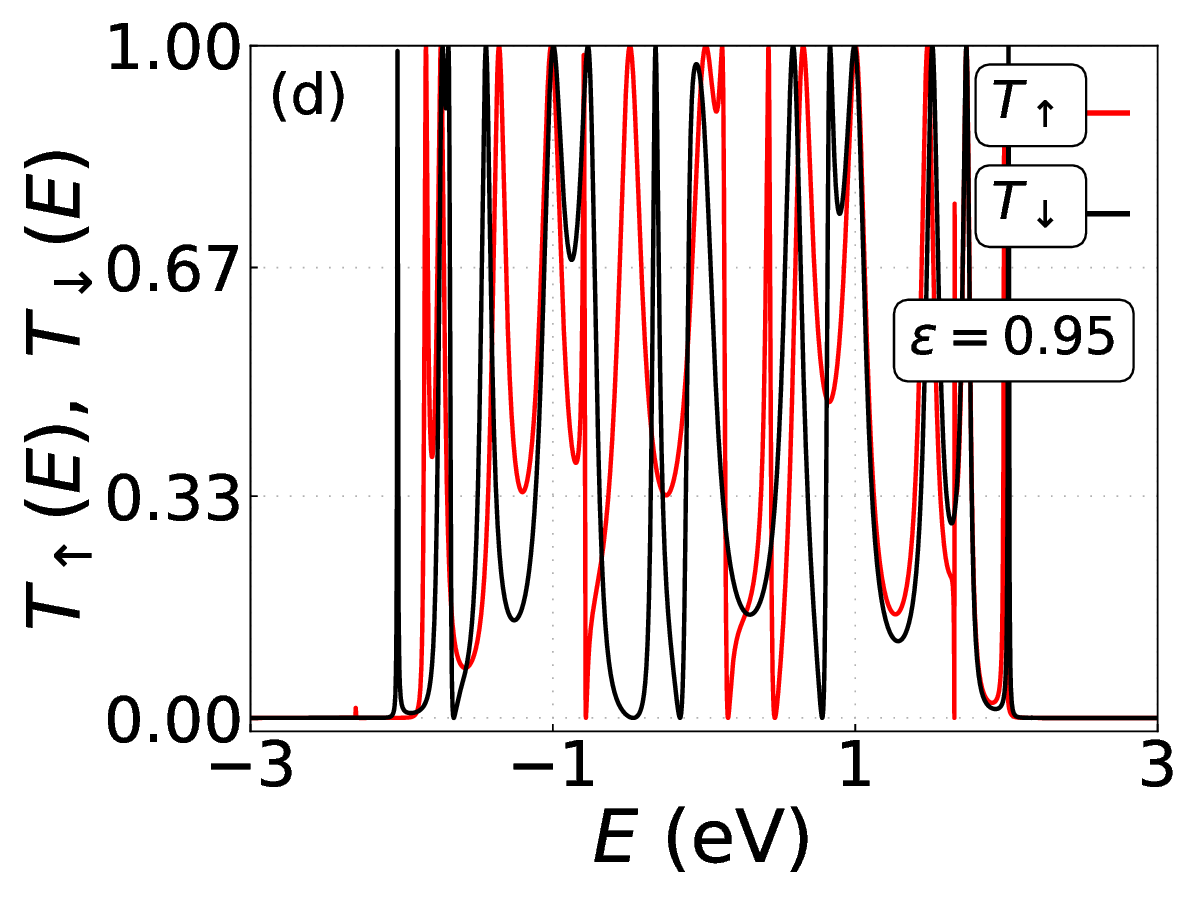}
\end{minipage}
\hfill
\begin{minipage}{0.32\textwidth}
\centering
\includegraphics[width=\linewidth,height=4.2cm]{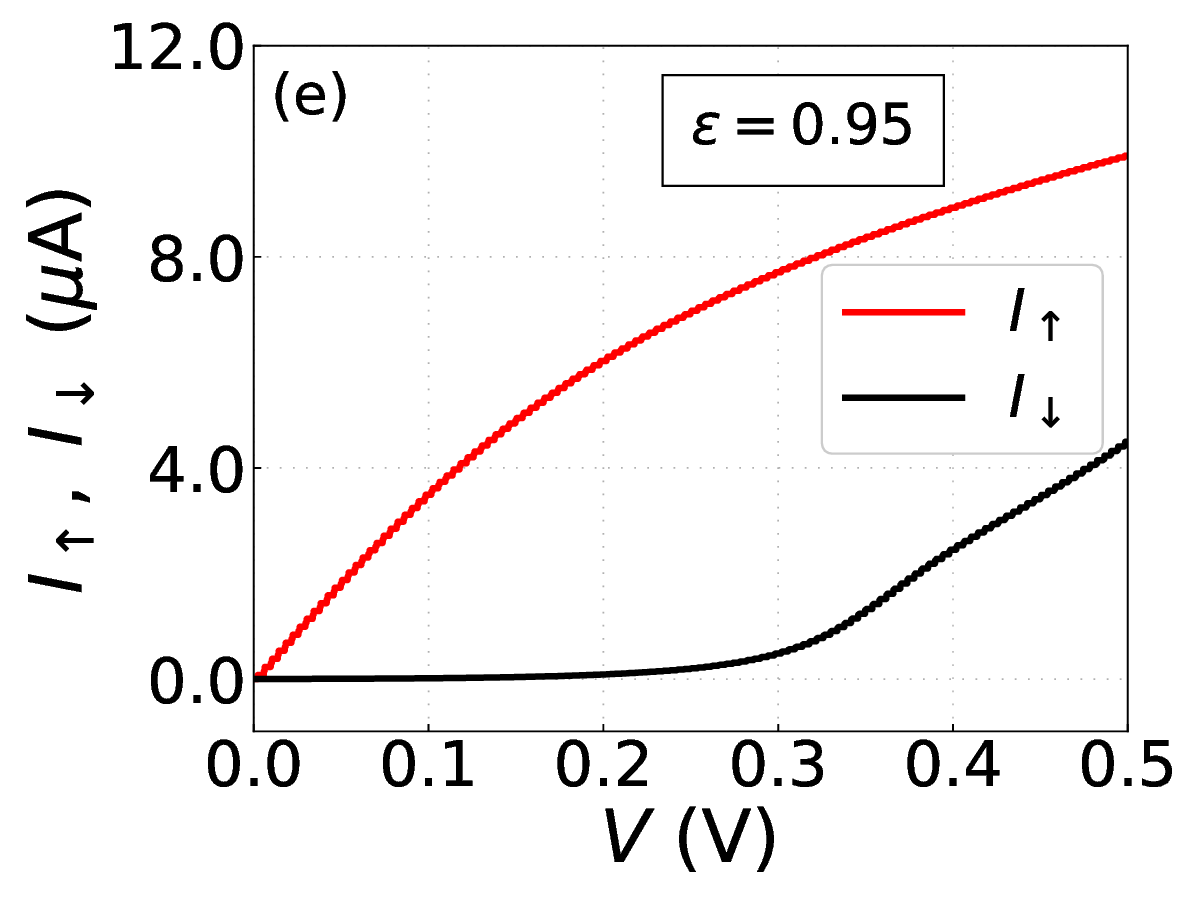}
\end{minipage}
\hfill
\begin{minipage}{0.32\textwidth}
\centering
\includegraphics[width=\linewidth,height=4.2cm]{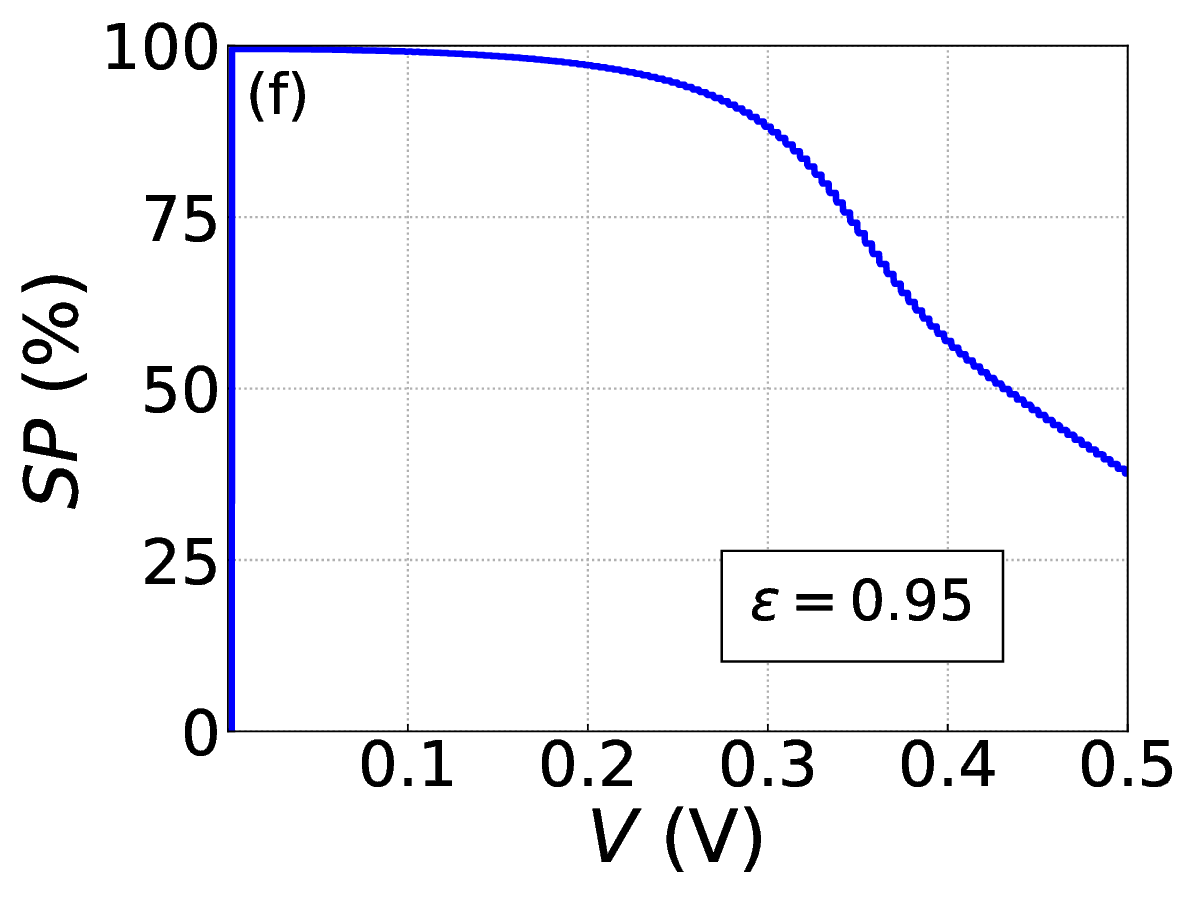}
\end{minipage}

\caption{(Color online). Spin-dependent transport characteristics of junction 1. Energy-dependent transmission spectra are presented in 
(a) and (d). The corresponding bias-voltage dependence of the spin-resolved current and spin polarization are shown in (b), (e) and (c), 
(f), respectively. Figures (a)-(c) correspond to $\varepsilon=0$, whereas figures (d)-(f) are obtained for $\varepsilon=0.95$.}
\label{fig:f2}
\end{figure*}

\section{Numerical results and discussion}

The numerical results for the proposed hybrid system are presented and discussed in this section. In particular, we analyze the effects of 
the in-plane electric field and ring-wire coupling on spin-dependent transport. The fixed model parameters used in the present calculations 
are specified below. The other parameters are varied accordingly and are mentioned at the appropriate places in the discussion. Throughout 
the calculations, we set the wire-to-electrode coupling strengths as $\tau_S=\tau_D=1\,$eV, the ring-wire coupling strength as $\gamma=1\,$eV,
and the electrode site energy as $\epsilon_0=0$. The spin-dependent scattering strength is taken as $h_n=h=0.7\,$eV. The NNH strength for 
the electrodes is fixed at $t_0=2\,$eV, whereas for the AFM ring and NM wire we set $t_r=t_w=1\,$eV. The site energy of the AFM ring in the
absence of an electric field, $\epsilon_n^{r(\varepsilon=0)}$, and the site energy of the wire, $\epsilon_n^w$, are set to zero.
In the first part of the discussion, 
we inspect the results for the junction 1 setup, while the second part is devoted to junction 2 configuration. Finally, a comparative 
analysis between the two junction configurations is presented, along with the robustness of the obtained results against variations in 
system size, temperature, and other relevant physical parameters associated with the model quantum system.

\subsection{Junction 1}

We begin the analysis with the transmission spectra of junction 1. The sizes of the AFM ring and NM wire are chosen as $N_r=12$ and 
$N_w=11$, respectively, unless specified otherwise. The ring is symmetrically side-attached to the wire, with the $1$st site of the AFM ring coupled to the $6$th 
site of the NM wire.
Figure~\ref{fig:f2} displays the spin-resolved transmission probability plotted against energy, along with the spin-dependent current 
and spin polarization with varying bias voltage. The top row corresponds to the field-free case ($\varepsilon=0$), while the bottom 
row corresponds to a finite in-plane electric field with $\varepsilon=0.95$.

In Fig.~\ref{fig:f2}(a), the spin-up and spin-down transmission spectra are shown for $\varepsilon=0$. Several transmission peaks appear 
\begin{figure}[htbp]
\centering\resizebox{8.5cm}{6cm}{\includegraphics{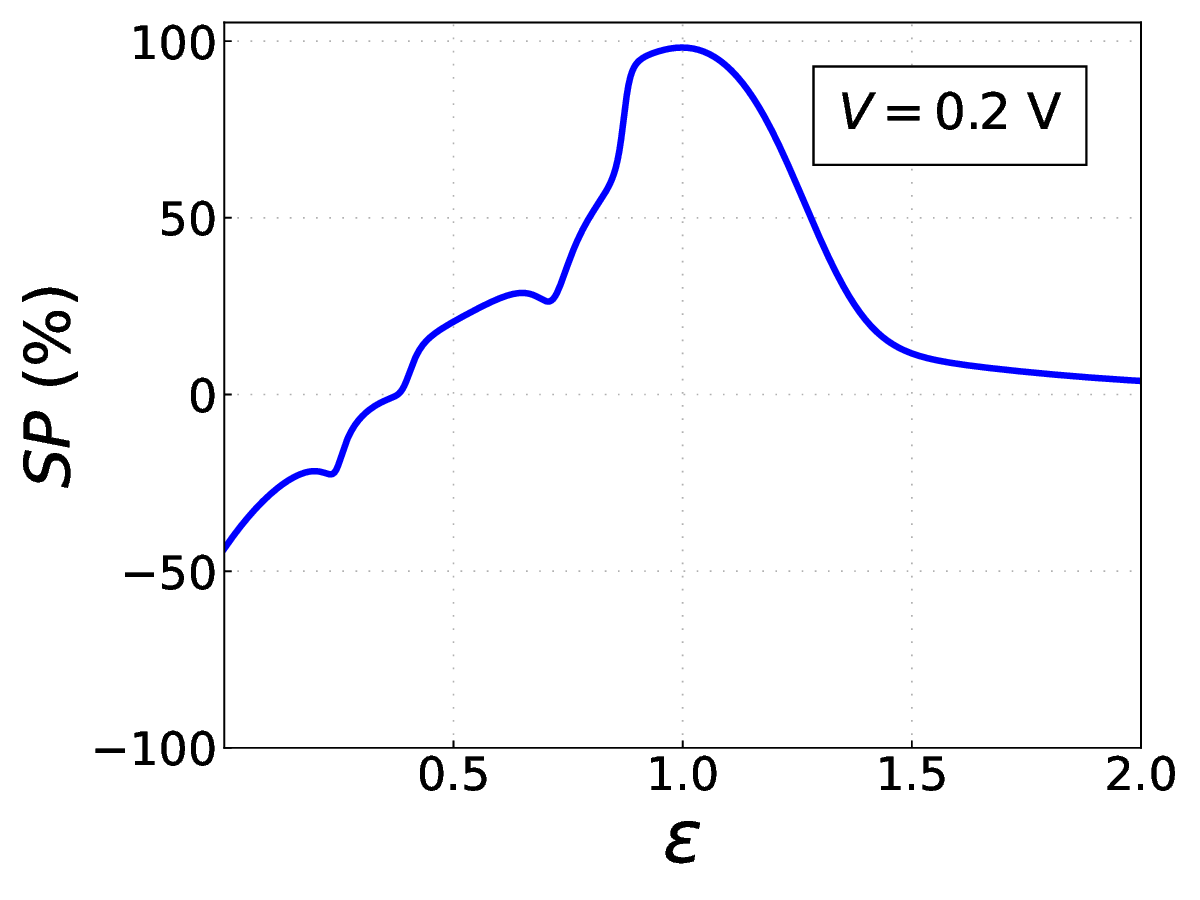}}
\caption{(Color online). Variation of spin polarization with the in-plane electric field at a constant bias voltage $V=0.2\,$V for 
the junction 1.} 
\label{fig:f3}
\end{figure}
within the energy window ($-3,3\,$eV), where each peak is associated with an eigenenergy of the ring-wire coupled system. 
For the present
impurity-free junction setup (junction 1), in the absence of an in-plane electric field, one would expect the symmetry between 
the up and down spin sub-Hamiltonians, $H_{rw,\uparrow}$ and $H_{rw,\downarrow}$, of the ring-wire coupled system to be preserved, resulting 
in completely overlapping transmission spectra. However, the figure exhibits a mismatch between the spin-resolved transmission probabilities.
Although the system is impurity-free and the AFM ring possesses zero net magnetization, symmetry breaking occurs due to the coupling of 
the wire to a site of the AFM ring. The underlying mechanism is as follows. Spin-dependent scattering exists only in the AFM ring, as 
the side-coupled wire is non-magnetic. In the absence of an external electric field, the staggered local magnetic moments along the spin
quantization axis lead to effective site energies for an up spin electron at different sites $\left\{n=1, 2, 3, \dots, M \right\}$ as 
$\left\{-h, +h, -h, +h, \dots \right\}$. In contrast, for a down spin electron, these site energies are $\left\{+h, -h, +h, -h, \dots 
\right\}$. The site energy sequence for down spin electrons is simply a single-site spatial translation of the up spin sequence. Under 
periodic boundary condition, this spatial translation maps the Hamiltonian of one spin species directly onto that of the other. Consequently,
owing to this spatial spin-sublattice symmetry, the energy eigenvalues for up and down spin electrons are exactly degenerate. This degeneracy 
is explicitly lifted when the AFM ring is coupled to the wire at a specific site, e.g., site $1$. The effect of the wire and its coupling 
to the ring can be incorporated as a renormalized site energy (say) $\Sigma$ at site $1$ of the ring. Hence, the effective site energies 
for up and down-spin electrons at site $1$ are $-h+\Sigma$ and $h+\Sigma$, respectively. Since the coupling modifies the local potentials 
for up and down spin electrons non-equivalently due to non-zero $h$, the symmetry is broken and the degeneracy is lifted. As a consequence 
of this broken degeneracy, spin-dependent transport emerges even in the \textit{absence} of an external field.

It is important to note that within the energy window ($-1,1\,$eV), the spin-resolved transmission spectra exhibit an approximately 
mirror-like behavior. A constructive transmission peak for one spin channel is accompanied by a suppressed transmission for the other, 
thereby reducing the overall mismatch between the two spectra. However, in the presence of the electric field, the 
mismatch between the 
spin up and spin down transmission spectra becomes more prominent, as shown in Fig.~\ref{fig:f2}(d). The application of an electric field
modifies the site energies of the AFM ring, making them site dependent and thereby introducing greater asymmetry between the up and 
down spin sub-Hamiltonians. This introduces a deterministic and smooth 
modulation of the ring site energies. Near the band edges, a few transmission peaks become suppressed due to the localization of the
corresponding electronic states induced by this non-uniform site-energy modulation. The applied electric field not only enlarges the 
overall transmission gap but also suppresses the mirror-like behavior between the spin-resolved transmission components.
\begin{figure*}[htbp]
	\centering
	\begin{minipage}{0.32\textwidth}
		\centering
		\includegraphics[width=\linewidth,height=4.2cm]{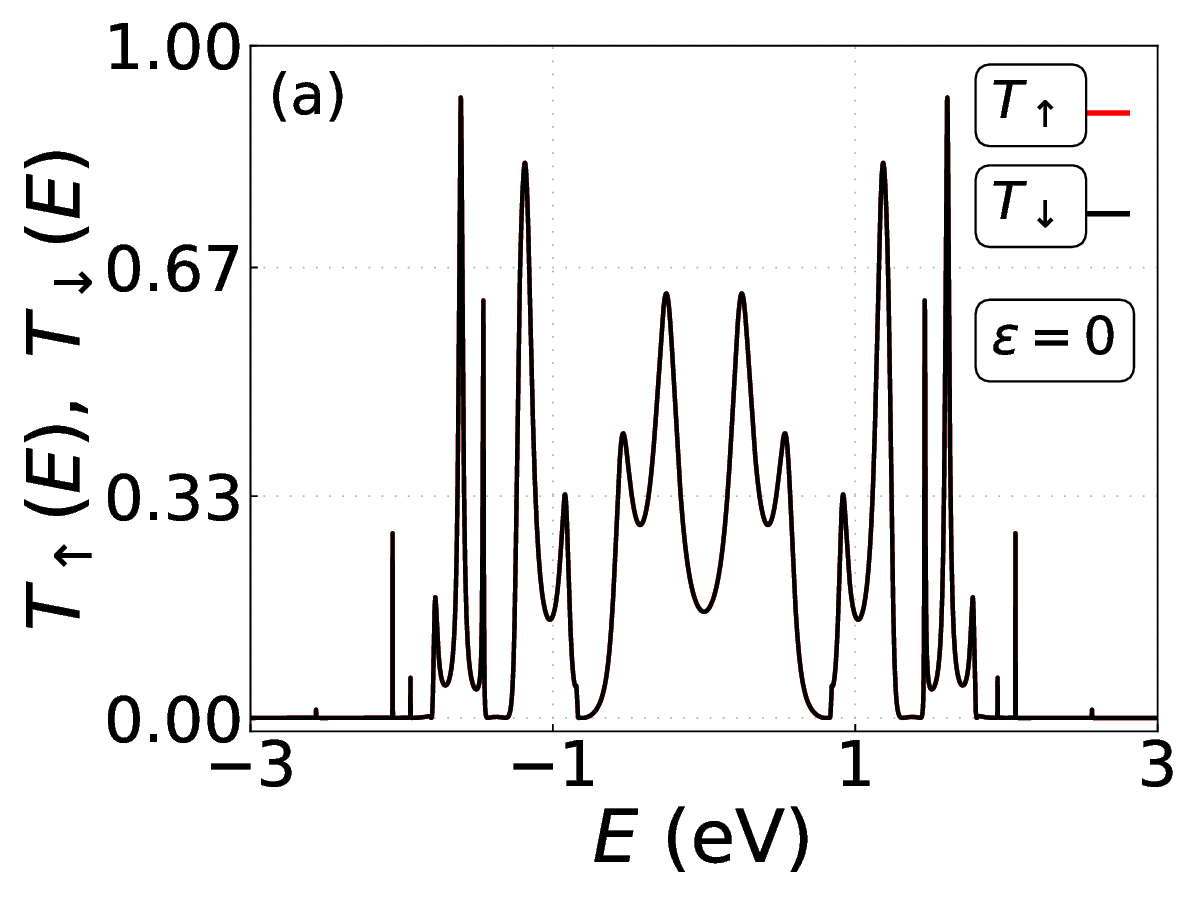}
	\end{minipage}
	\hfill
	\begin{minipage}{0.32\textwidth}
		\centering
		\includegraphics[width=\linewidth,height=4.2cm]{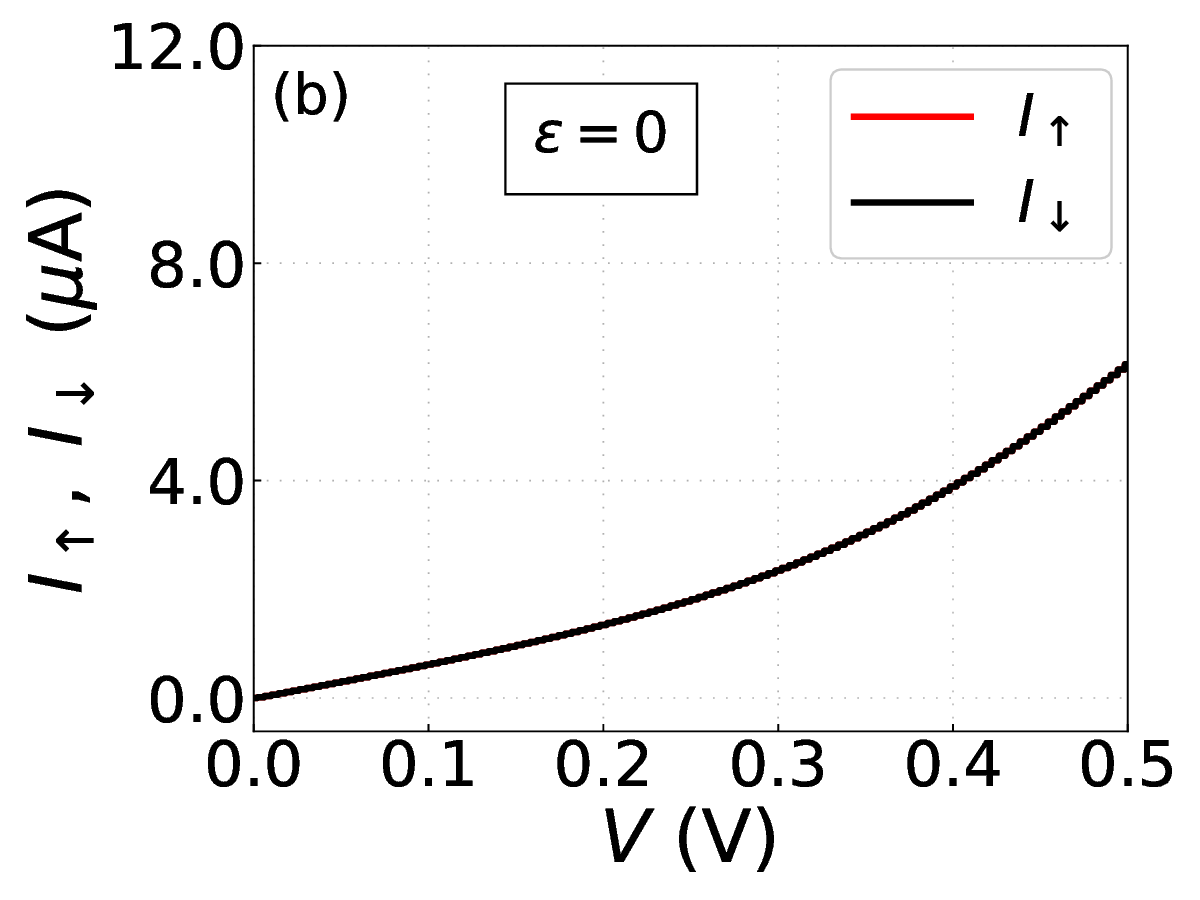}
	\end{minipage}
	\hfill
	\begin{minipage}{0.32\textwidth}
		\centering
		\includegraphics[width=\linewidth,height=4.2cm]{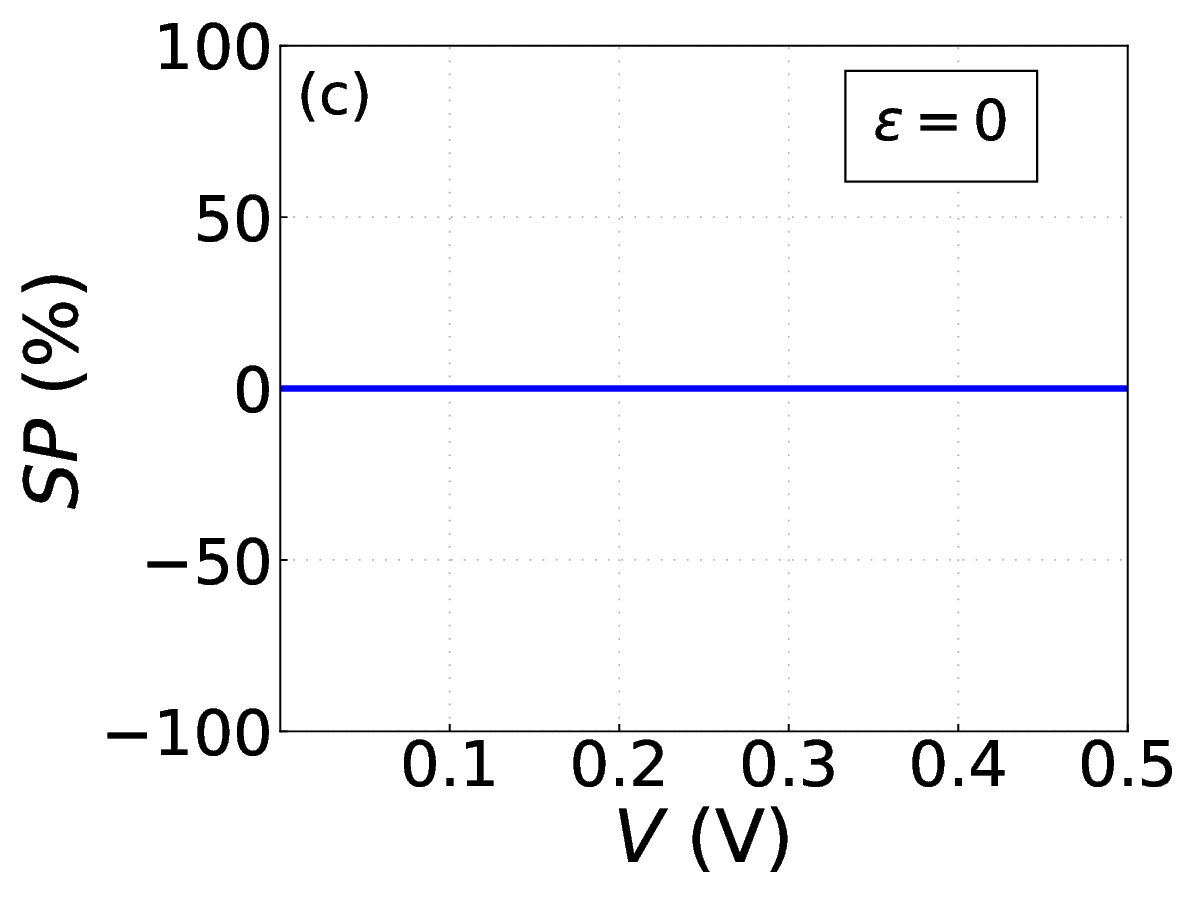}
	\end{minipage}
	
	\vspace{0.2cm}
	
	\begin{minipage}{0.32\textwidth}
		\centering
		\includegraphics[width=\linewidth,height=4.2cm]{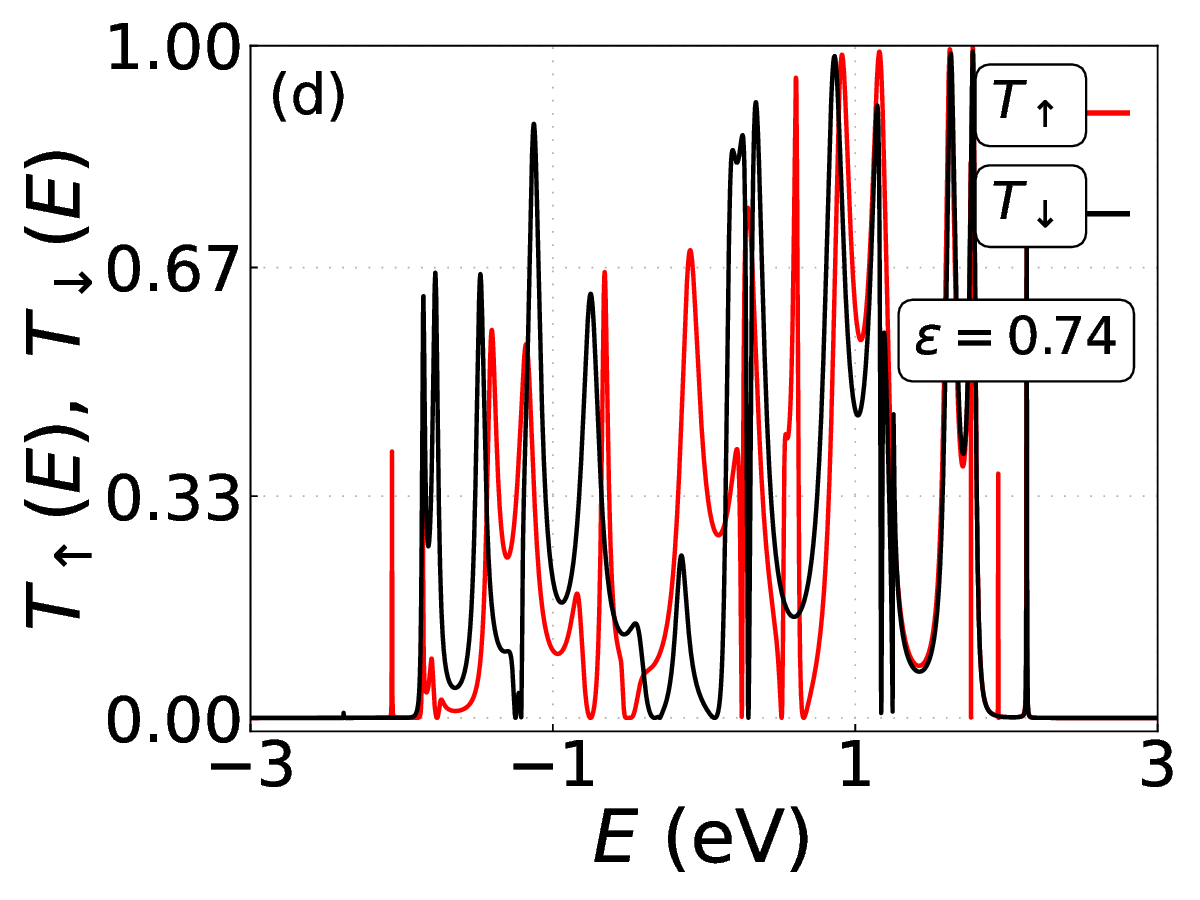}
	\end{minipage}
	\hfill
	\begin{minipage}{0.32\textwidth}
		\centering
		\includegraphics[width=\linewidth,height=4.2cm]{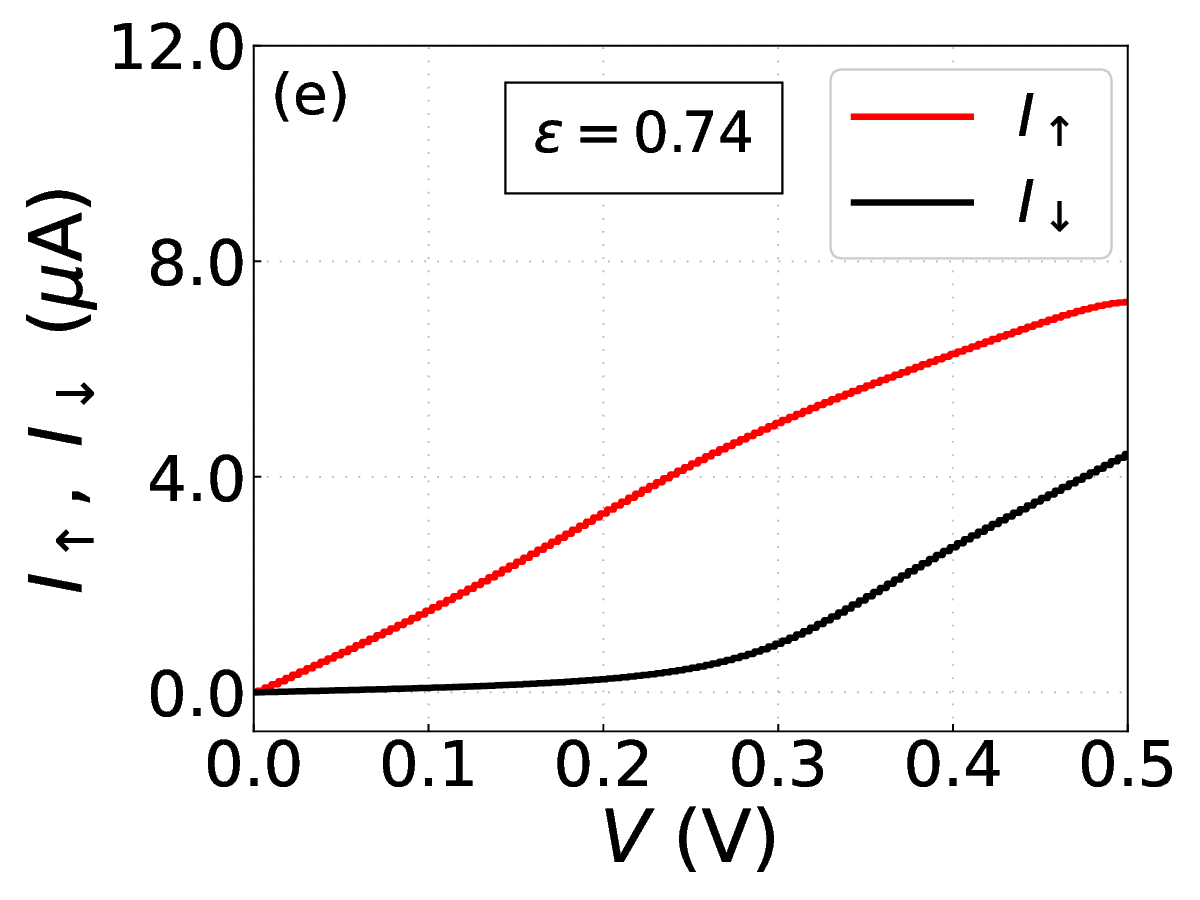}
	\end{minipage}
	\hfill
	\begin{minipage}{0.32\textwidth}
		\centering
		\includegraphics[width=\linewidth,height=4.2cm]{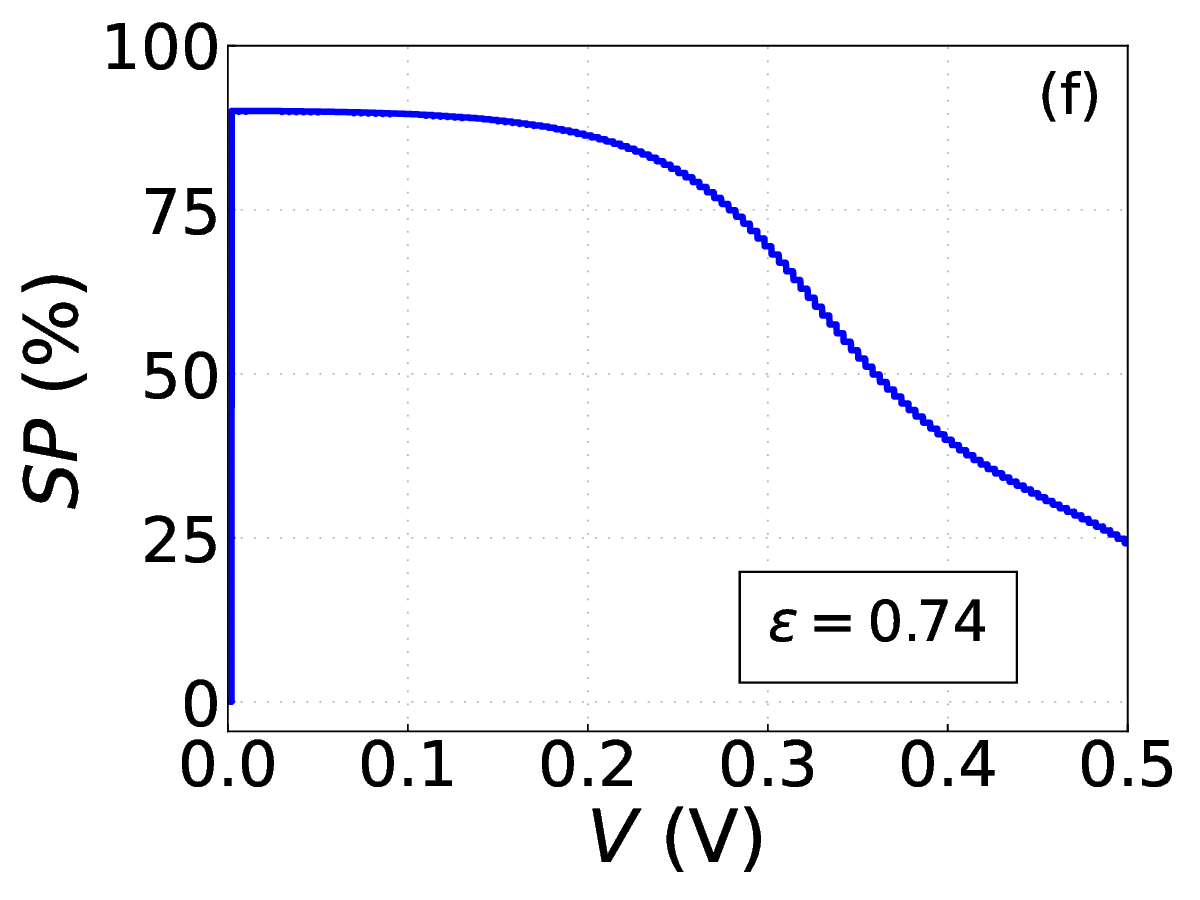}
	\end{minipage}
	\caption{(Color online). Spin-dependent transport characteristics of junction 2. The energy dependence of the transmission probability is displayed in (a) and (d). The corresponding spin-dependent current and spin polarization as functions of bias voltage are shown in (b), (e) and (c), (f), respectively. Figures (a)--(c) correspond to $\varepsilon=0$, while figures (d)--(f) represent the case $\varepsilon=0.74$.}
	\label{fig:f4}
\end{figure*}

The spin-dependent currents at $E_F=-0.5\,$eV, shown in Figs.~\ref{fig:f2}(b) and \ref{fig:f2}(e), follow the corresponding spin-resolved transmission spectra closely. A noticeable gap between the spin-up and spin-down current appears even at $\varepsilon=0$. In the presence of a finite in-plane electric field (Fig.~\ref{fig:f2}(e)), the down-spin current vanishes up to a certain bias voltage while the up-spin current remains considerably large, resulting in a strongly spin-polarized current over a moderate bias window.

The spin polarization is illustrated in Figs.~\ref{fig:f2}(c) and \ref{fig:f2}(f), corresponding to $\varepsilon=0$ and $\varepsilon=0.95$, respectively. In the absence of an in-plane electric field, the SP remains finite but is not sufficiently large, owing to the partial 
symmetry breaking induced by the single-coupling. In contrast, a finite in-plane electric field drives the SP to nearly $100\%$ already in the low-bias region, maintaining this high degree of polarization over a wide bias window. This behavior arises from the complete suppression of one spin-current component within this bias range.

To investigate the role of the in-plane electric field, we consider the variation of SP with $\varepsilon$ at a constant bias voltage 
$V=0.2,$V, as an illustrative example, as shown in Fig.~\ref{fig:f3}. The SP increases with increasing field strength and reaches a maximum
value at a particular field strength. Beyond this point, it gradually decreases and eventually tends toward zero for sufficiently large 
field strengths. The application of the in-plane electric field modifies the site energies of the AFM ring, and the resulting
site energy modulation becomes stronger with increasing field strength. This initially enhances the spin-selective transport, leading to an
increase in SP. However, beyond an optimal field strength, the stronger non-uniform site-energy modulation progressively suppresses the
contribution of the ring states to the transport process. At sufficiently large $\varepsilon$, the strong site-energy modulation increasingly
suppresses the participation of the AFM ring in transport, causing the ring to gradually decouple from the NM wire. Consequently, the
spin-filtering effect is progressively weakened, and the SP tends toward zero at higher field strengths. It is also noted that at
$\varepsilon=0$, a finite SP is observed due to the symmetry breaking introduced by the single coupling, consistent with the results 
discussed above.

\subsection{Junction 2}

We now present the numerical results for junction 2 and analyze the corresponding features. 
\begin{figure}[htbp]
\centering
\resizebox{8cm}{5cm}{\includegraphics{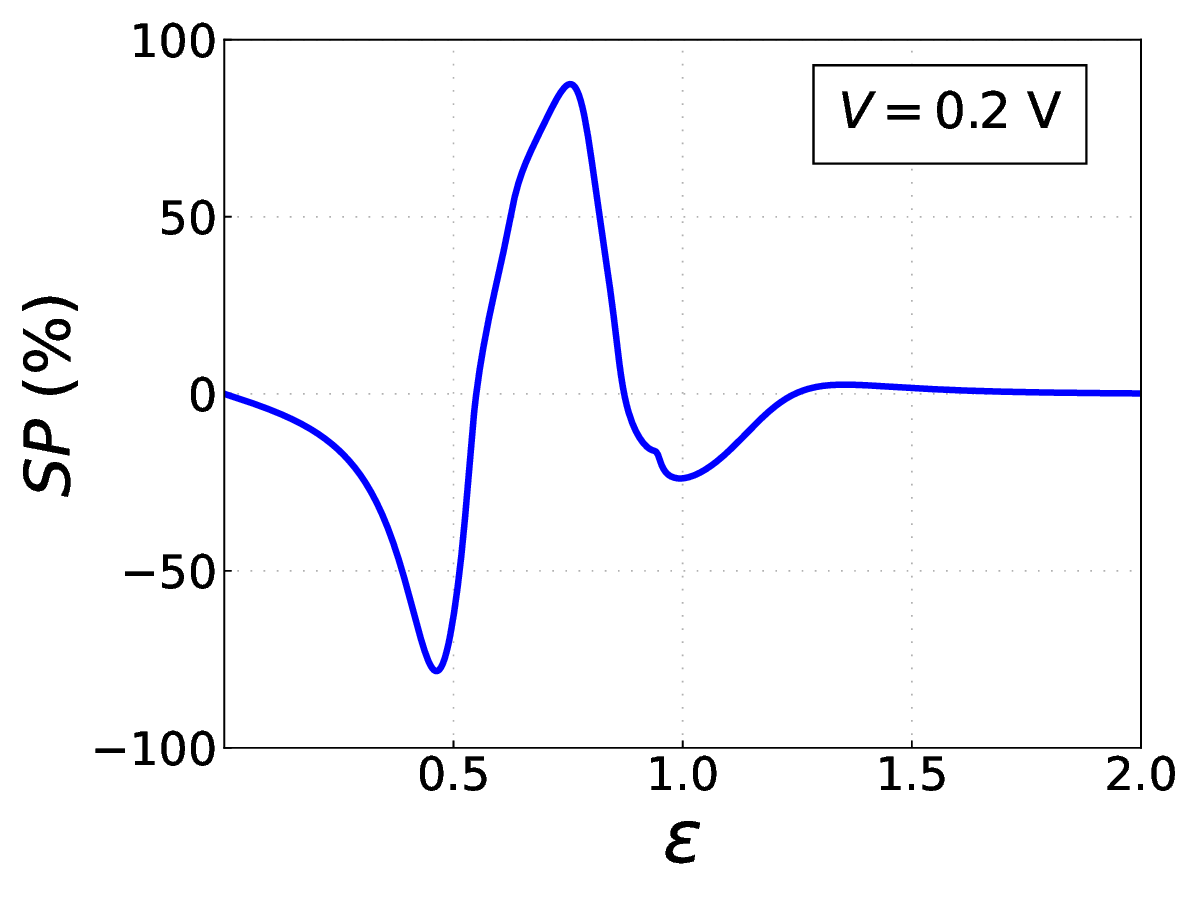}}
\caption{(Color online). Variation of spin polarization with the in-plane electric field at a constant bias voltage $V=0.2\,$V 
for junction 2.}
\label{fig:f5}
\end{figure}
The sizes of the AFM ring and NM wire are chosen as $N_r=14$ and $N_w=12$, respectively, unless mentioned otherwise.

The ring is symmetrically side-attached to the wire, with the $1$st and $2$nd sites of the AFM ring coupled to the $6$th and $7$th sites 
of the NM wire through two adjacent equal-strength couplings ($\gamma=1\,$eV).
Figure~\ref{fig:f4} presents the same set of spin-dependent quantities as Fig.~\ref{fig:f2}, namely the transmission probability, current, and spin polarization, but now for junction 2, with $\varepsilon=0$ (first row) and $\varepsilon=0.74$ (second row).
\begin{figure}[htbp]
\centering
\includegraphics[width=0.94\columnwidth]{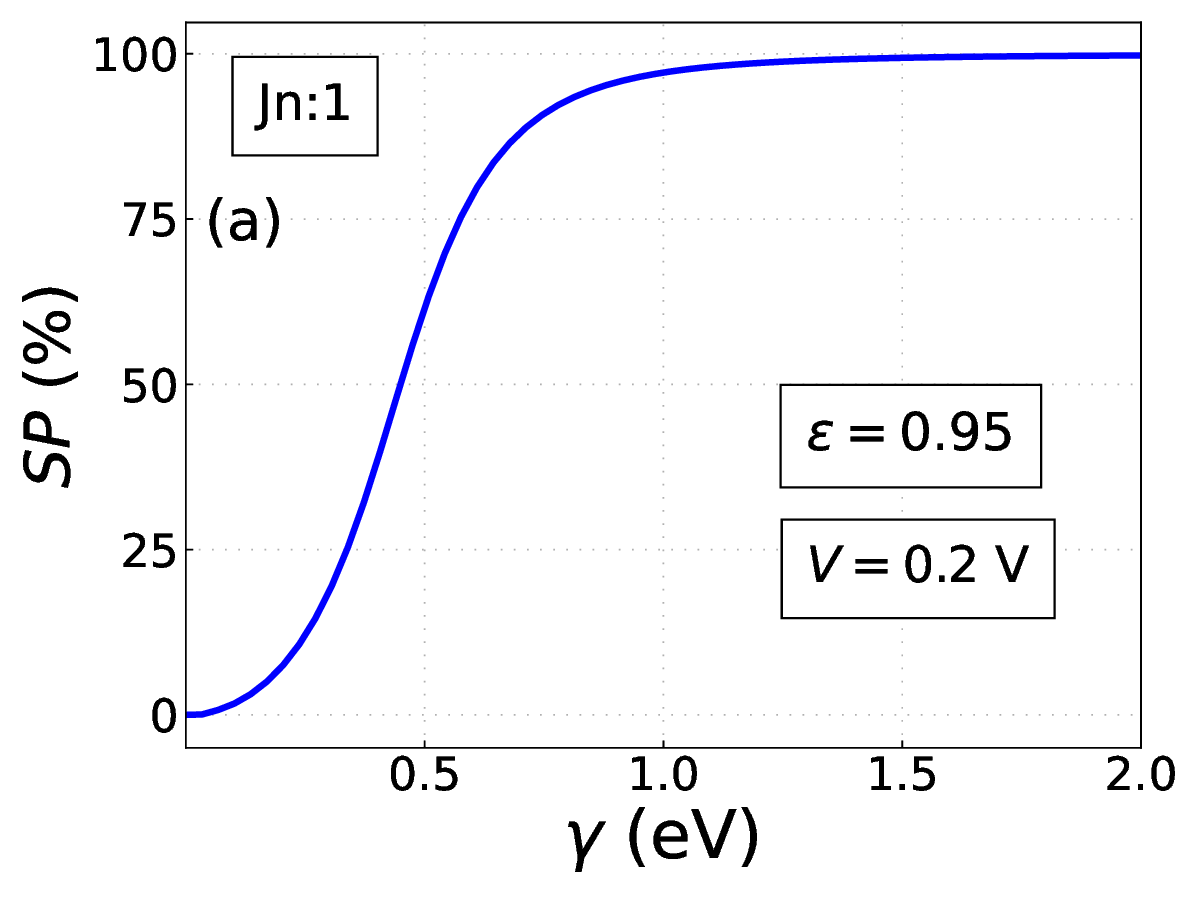}
	
\vspace{0.3cm}
	
\includegraphics[width=0.95\columnwidth]{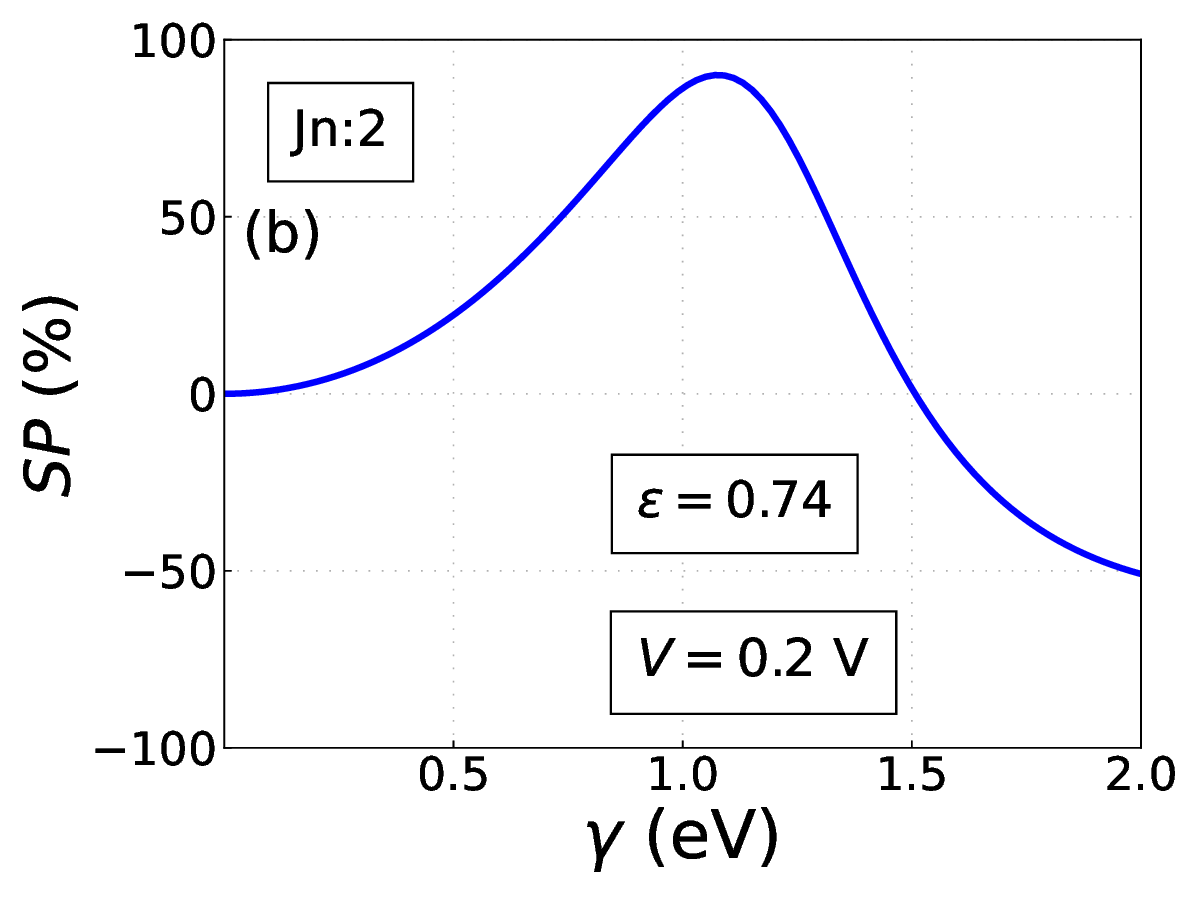}
\caption{(Color online). The variation of spin polarization with the ring-wire coupling strength at a constant bias voltage $V=0.2\,$V 
for (a) junction 1 with $\varepsilon=0.95$ and (b) junction 2 with $\varepsilon=0.74$.}
\label{fig:f6}
\end{figure}

In the absence of an in-plane electric field, the energy dependence of the transmission probability is presented in Fig.~\ref{fig:f4}(a). 
The spin-resolved transmission components overlap completely, as expected, which is markedly different from the behavior observed in 
junction 1. In junction 1, the symmetry between $H_{rw,\uparrow}$ and $H_{rw,\downarrow}$ is broken due to the single-site coupling. 
Following this argument, one might also expect a mismatch between the up and down spin transmission probabilities in the present case. 
However, unlike the single-coupling configuration, where electrons interact predominantly with a specific local magnetic moment, the
double-coupling arrangement allows electrons to interact symmetrically with both positive and negative staggered magnetic moments. 
Consequently, the effective spin symmetry of the system is restored, resulting in identical transport behavior for up and down spin 
electrons. The existence of spin symmetry in the double-coupling arrangement can be understood from the following argument. In the AFM 
ring, the two contact sites possess magnetic moments oriented in opposite directions. With symmetric dual-bond connections, the effective 
site energies at these two sites for an up spin electron are $-h+\Sigma$ and $+h+\Sigma$, while those for a down spin electron are 
$+h+\Sigma$ and $-h+\Sigma$. The uncoupled ring sites retain their standard alternating values ($\pm h$). It is therefore evident that 
the effective site energies at site numbers $1$ and $2$ are interchanged under the sub-lattice exchange operation when the spin is 
reversed. Hence, the energy eigenvalues remain degenerate for the two spin cases, resulting in zero spin polarization in the field-free case.

The symmetry is broken once the electric field is applied, as the site energies of the AFM ring become site dependent and the corresponding
effective site-energy configurations for up and down spin electrons are no longer related by the sub-lattice exchange symmetry. We find a
significant mismatch between the spin-resolved transmission components in the non-zero-field case, as shown in Fig.~\ref{fig:f4}(d). Thus, 
in the double-coupling configuration, the in-plane electric field provides the origin of the symmetry breaking and consequently gives rise 
to spin-dependent transport.

The spin-dependent currents as a function of bias voltage are shown in Fig.~\ref{fig:f4}(b) for $\varepsilon=0$
\begin{figure}[htbp]
	\centering
	\includegraphics[width=0.95\columnwidth]{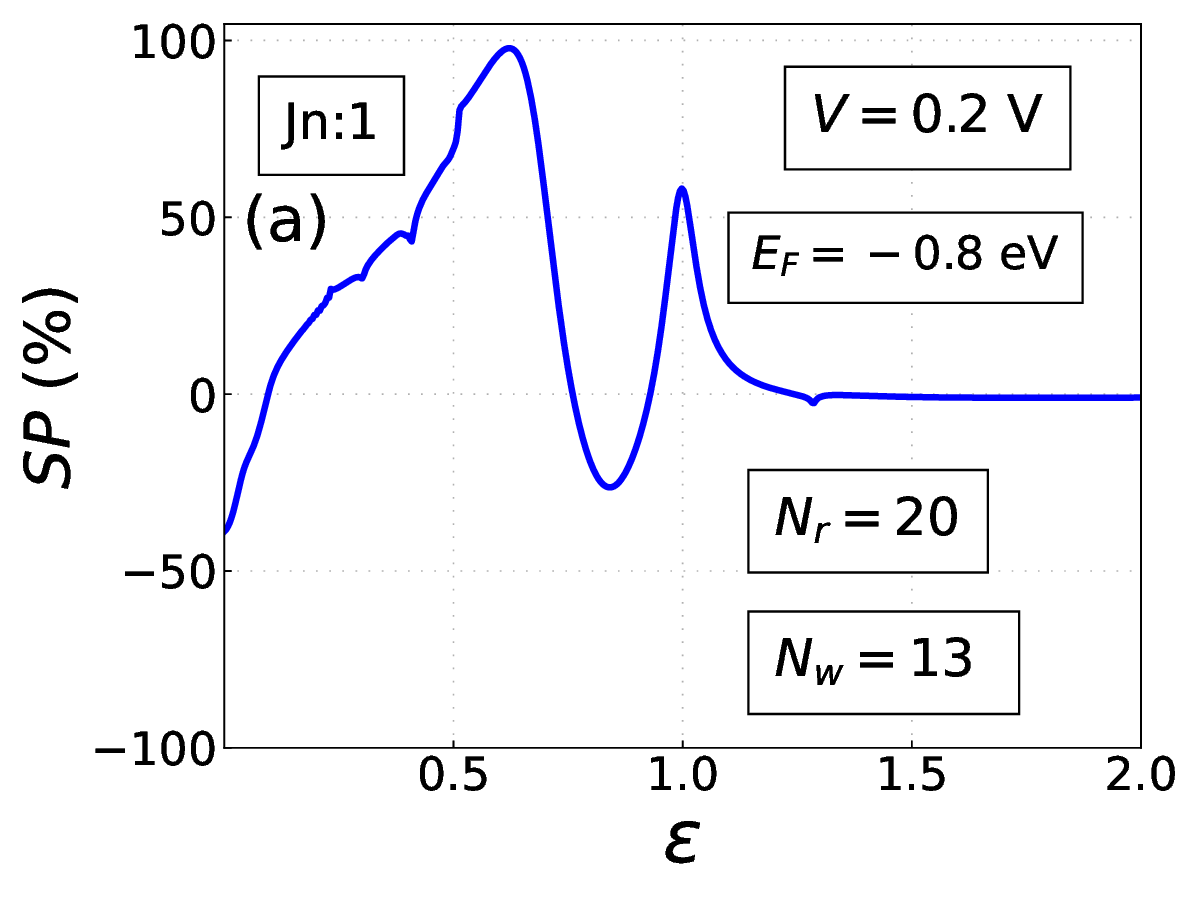}
	
	\vspace{0.3cm}
	
\includegraphics[width=0.95\columnwidth]{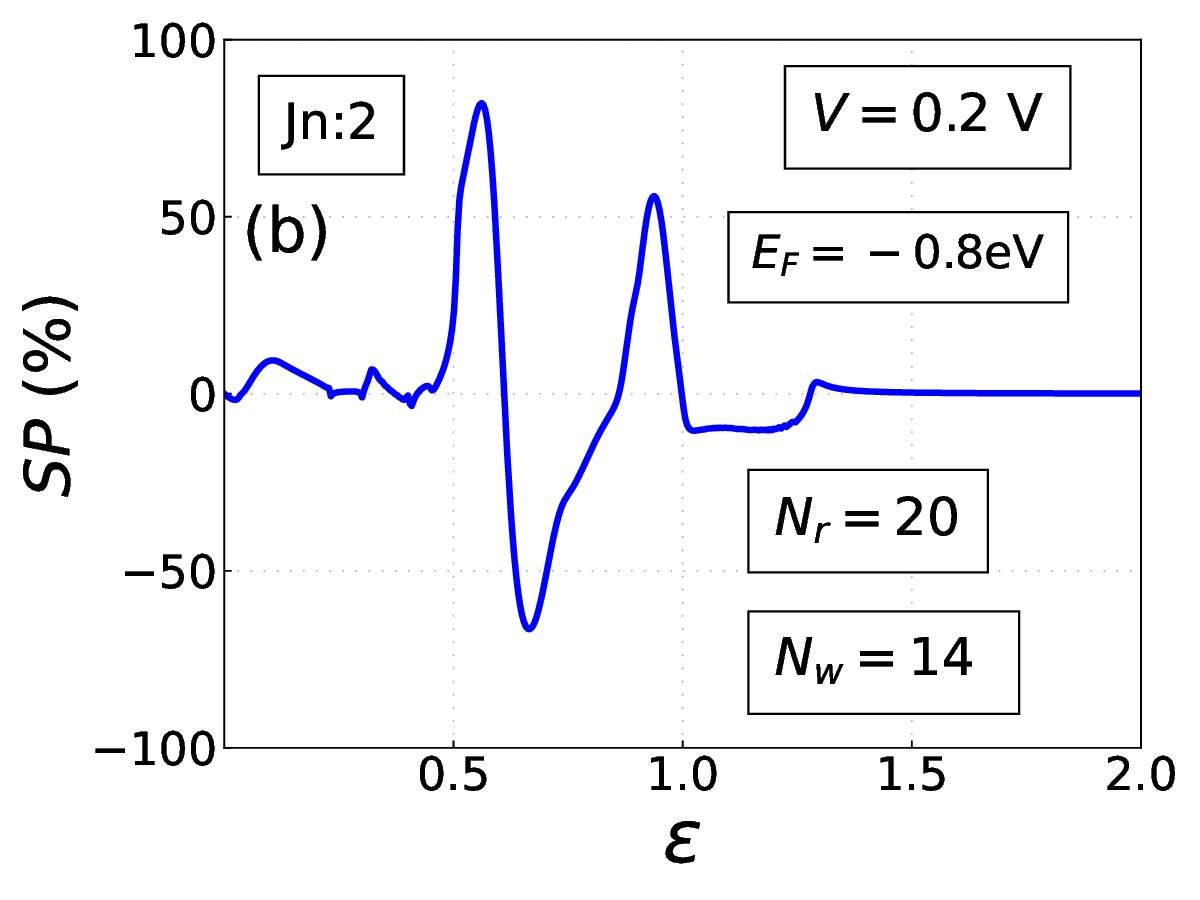}
\caption{(Color online). Spin polarization as a function of the in-plane electric field for different system sizes at a fixed bias voltage
$V=0.2\,$V. Here, the system sizes are chosen differently from the earlier results to examine the robustness of the obtained features. 
Figures (a) and (b) correspond to junction 1 with $(N_r,N_w)=(20,13)$ and junction 2 with $(N_r,N_w)=(20,14)$, respectively.}
\label{fig:f7}
\end{figure}
and Fig.~\ref{fig:f4}(e) for $\varepsilon=0.74$ at the Fermi energy $E_F=0$. In the absence of an in-plane electric field, the up and down
spin currents become degenerate, consistent with the restored spin symmetry in the double-coupling configuration. This behavior is in sharp
contrast to junction 1, where the single-coupling breaks the spin symmetry. In the presence of a finite in-plane electric field, the
spin-resolved current components become well separated, as illustrated in Fig.~\ref{fig:f4}(e). This behavior originates from the 
field-induced symmetry breaking discussed above.
\begin{figure*}[htbp]
	\centering
	\begin{minipage}{0.32\textwidth}
		\centering
		\includegraphics[width=\linewidth,height=4.2cm]{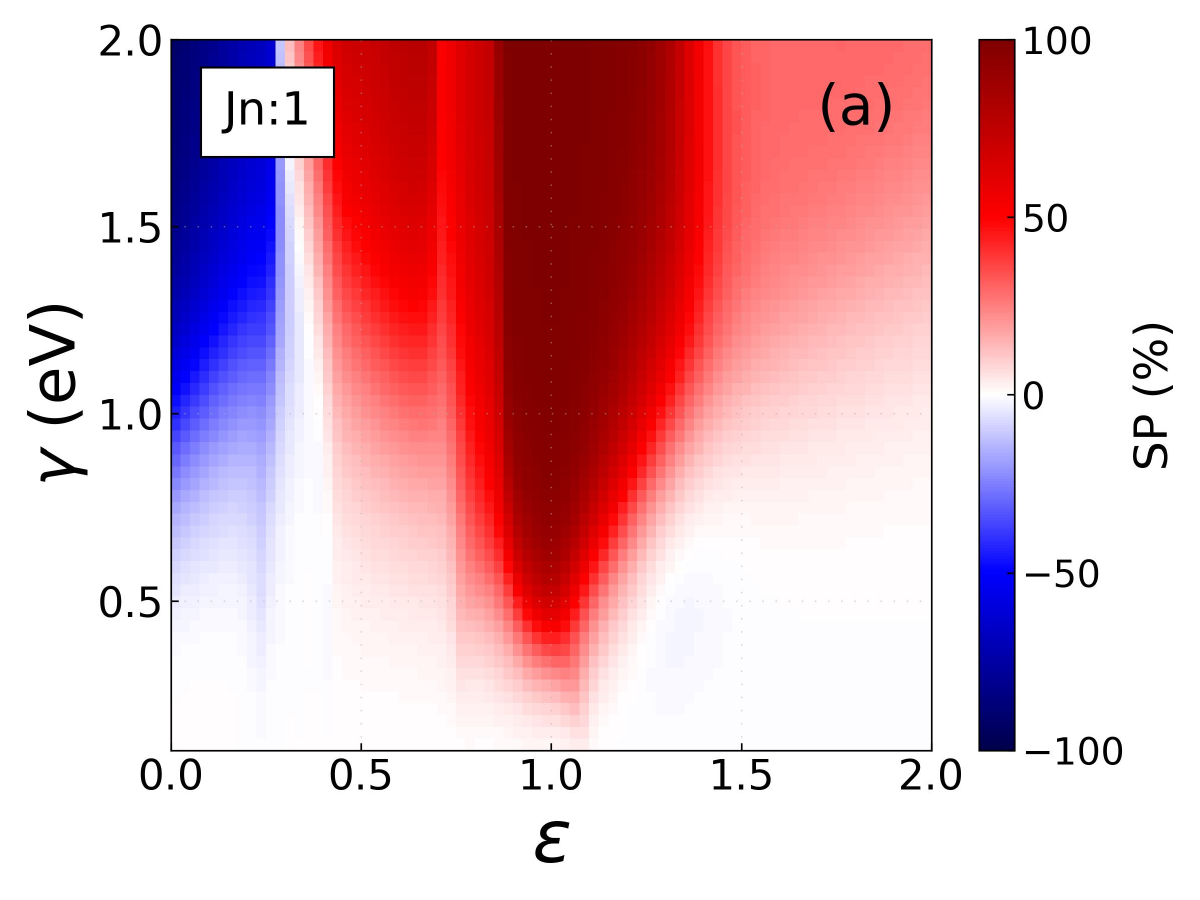}
	\end{minipage}
	\hfill
	\begin{minipage}{0.32\textwidth}
		\centering
		\includegraphics[width=\linewidth,height=4.2cm]{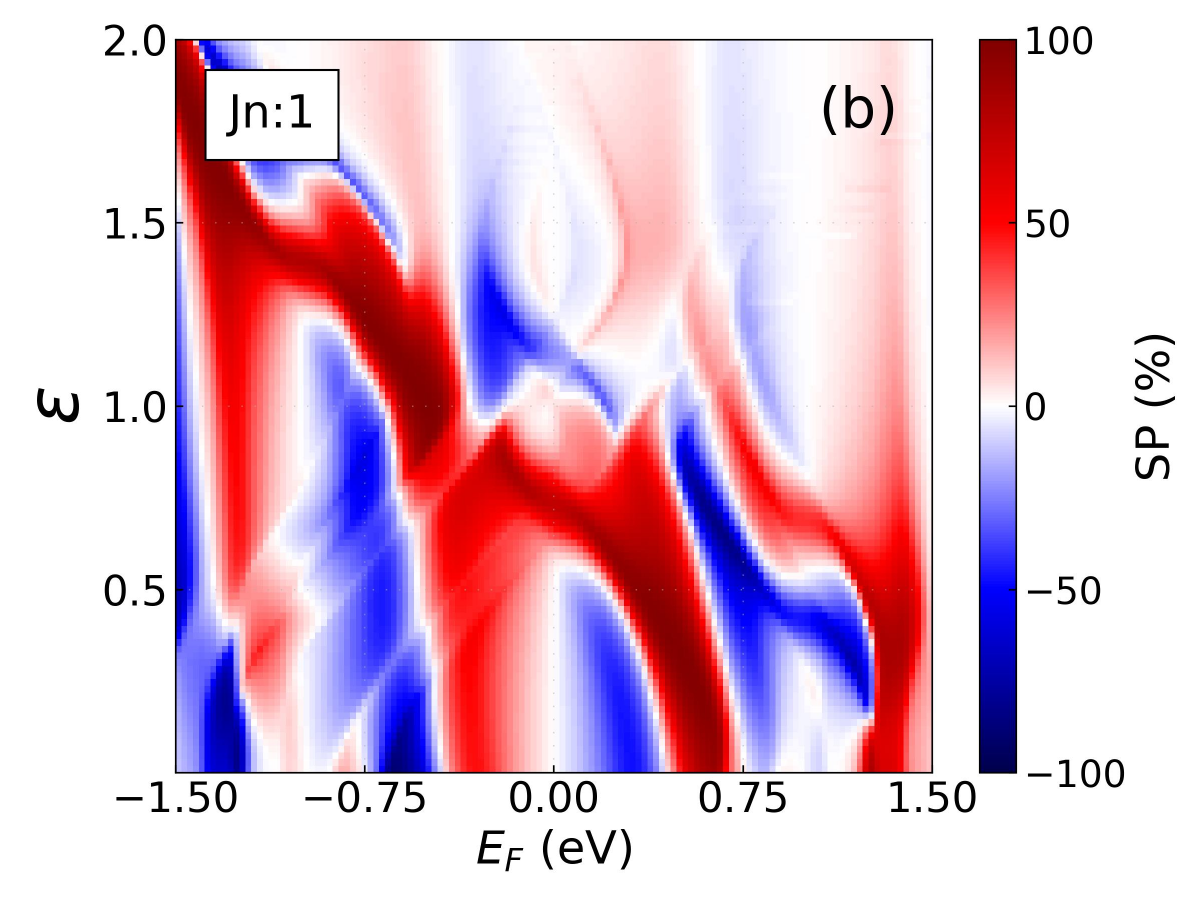}
	\end{minipage}
	\hfill
	\begin{minipage}{0.32\textwidth}
		\centering
		\includegraphics[width=\linewidth,height=4.2cm]{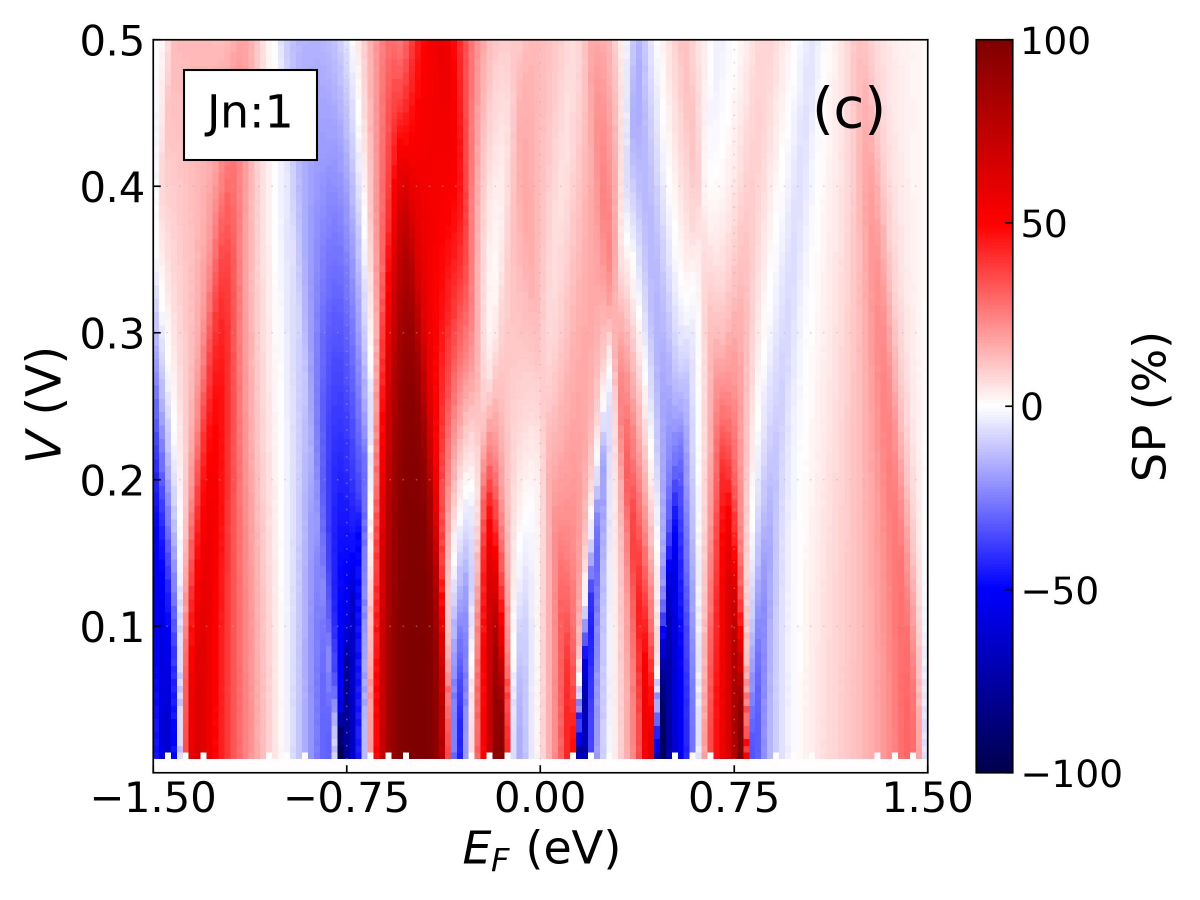}
	\end{minipage}
	
	\vspace{0.2cm}
	
	\begin{minipage}{0.32\textwidth}
		\centering
		\includegraphics[width=\linewidth,height=4.2cm]{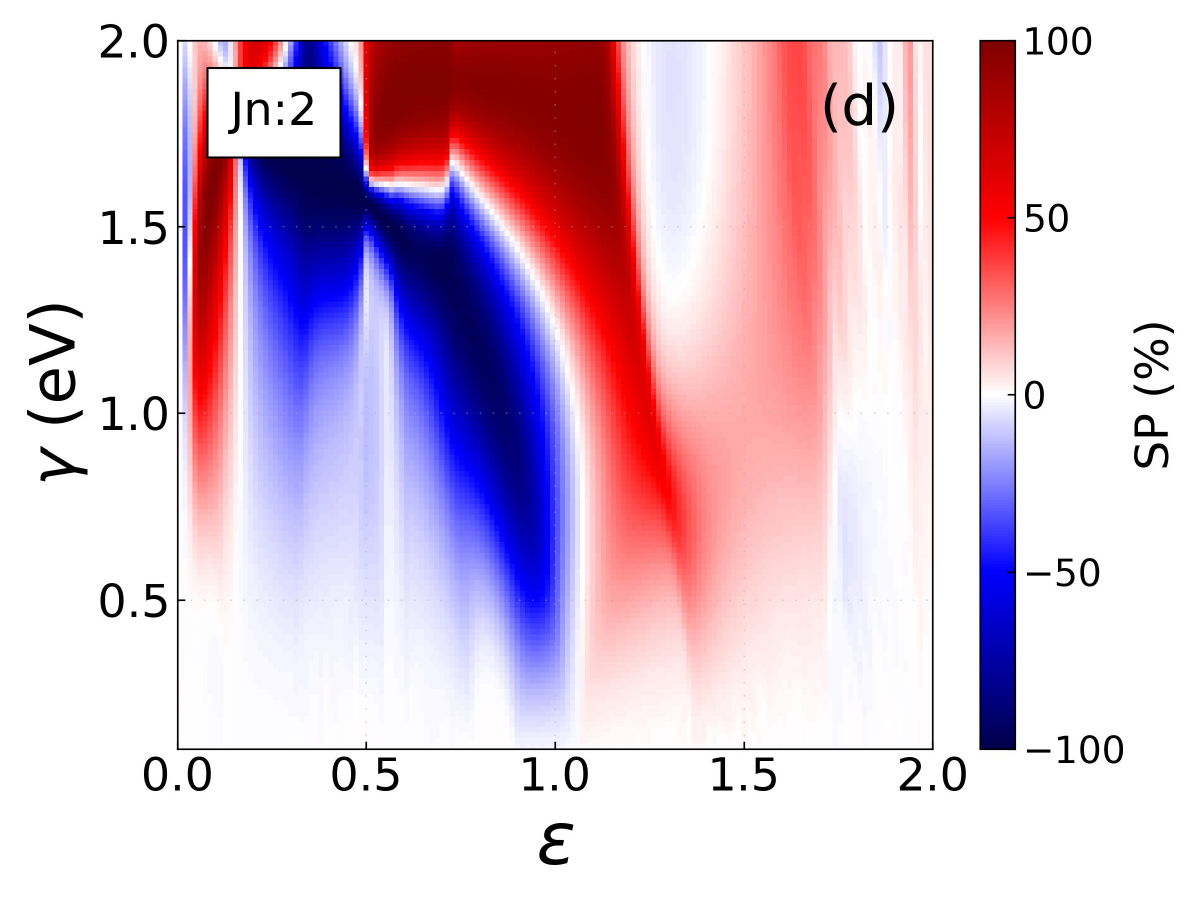}
	\end{minipage}
	\hfill
	\begin{minipage}{0.32\textwidth}
		\centering
		\includegraphics[width=\linewidth,height=4.2cm]{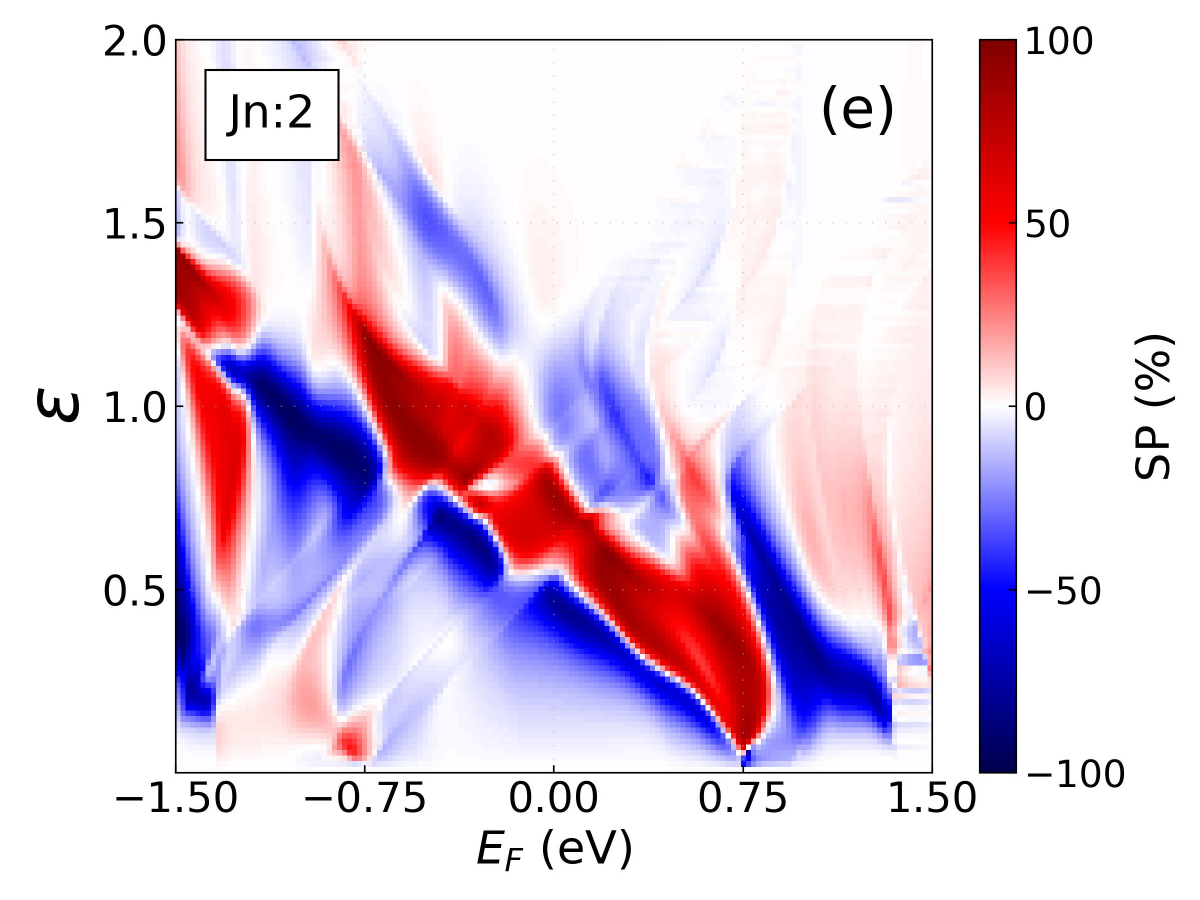}
	\end{minipage}
	\hfill
	\begin{minipage}{0.32\textwidth}
		\centering
		\includegraphics[width=\linewidth,height=4.2cm]{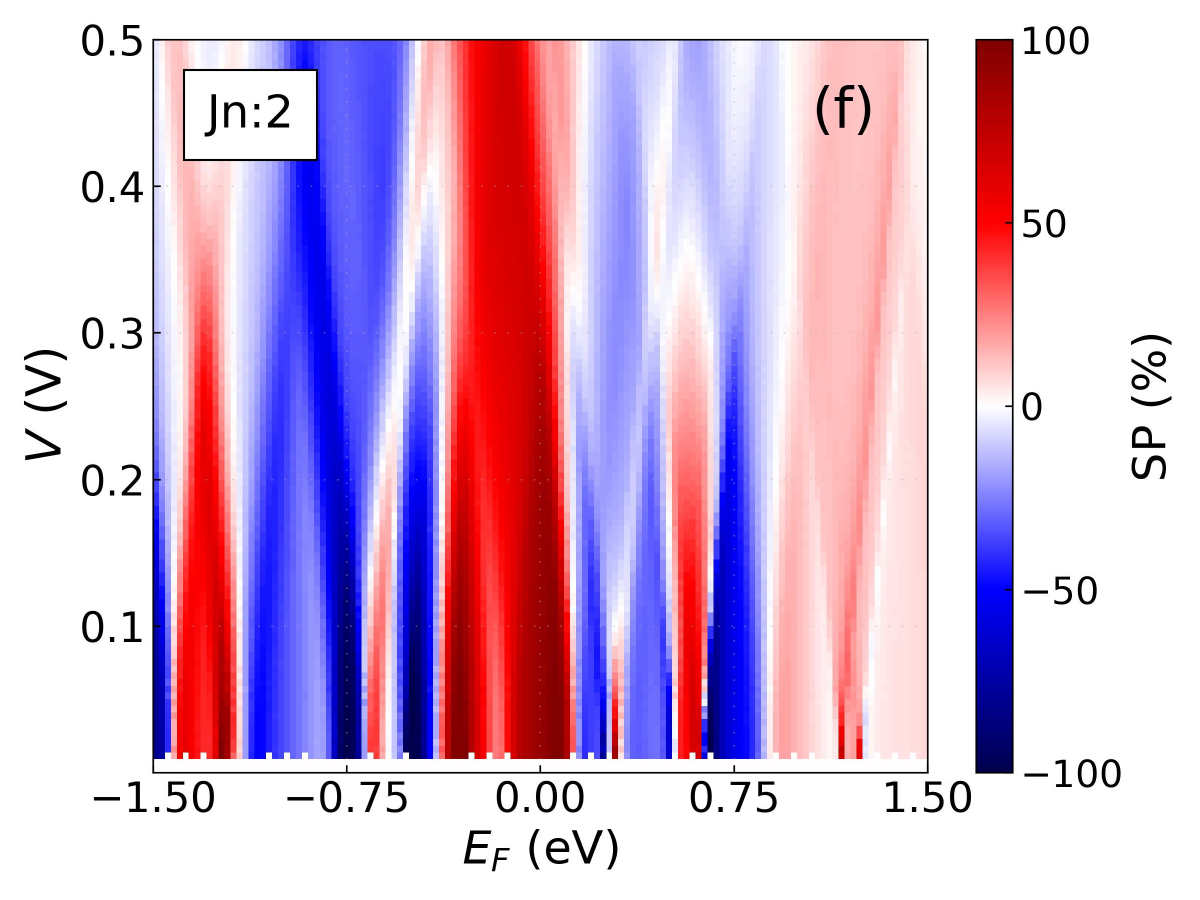}
	\end{minipage}
	\caption{(Color online). Density plots of spin polarization for different parameter regimes in junction 1 and junction 2. The variation of SP with ring-wire coupling and electric field is shown in (a) and (d), with Fermi energy and electric field in (b) and (e), and with Fermi energy and bias voltage in (c) and (f). Figures (a)--(c) correspond to junction 1, while figures (d)--(f) represent junction 2.}
	\label{fig:density}
\end{figure*}
In the adjacent double-coupling configuration (junction 2), all spin symmetries remain preserved in the absence of an in-plane electric field, resulting in zero spin polarization, as shown in Fig.~\ref{fig:f4}(c). In contrast, upon applying a finite in-plane electric field, a significantly large spin polarization (above $80\%$) is achieved, as illustrated in Fig.~\ref{fig:f4}(f).

The SP as a function of $\varepsilon$ at $V=0.2\,$V, as an illustrative example, is shown in Fig.~\ref{fig:f5}. At $\varepsilon=0$, the 
SP is zero, consistent with the degenerate spin currents discussed above. With increasing field strength, spin asymmetry is introduced 
into the AFM ring, causing the down-spin current to dominate up to $\varepsilon\approx0.5$, yielding a large negative SP exceeding $-90\%$.
Beyond this region, the up-spin current becomes dominant up to $\varepsilon\approx0.8$, resulting in a high positive SP of $\sim100\%$. 
At sufficiently large field strengths, the AFM ring gradually decouples from the NM wire, and the SP tends toward zero. These findings 
suggest that the in-plane electric field serves as an effective knob for controlling the spin polarization.

To further compare the transport characteristics of the two junctions, the SP as a function of the ring-wire coupling strength $\gamma$ 
at $V=0.2\,$V is shown in Figs.~\ref{fig:f6}(a) and \ref{fig:f6}(b) for junction 1 and junction 2, respectively.
For junction 1, the SP gradually increases with increasing ring-wire coupling strength and eventually saturates at sufficiently large 
$\gamma$, as illustrated in Fig.~\ref{fig:f6}(a). Here, the electric field strength is fixed at $\varepsilon=0.95$. At $\gamma=0$, the 
AFM ring remains completely decoupled from the NM wire, preventing the electrons in the wire from interacting with the field-modulated
spin-asymmetric ring states, and consequently the SP becomes zero. As $\gamma$ increases, the interaction between the wire electrons 
and the spin-dependent states of the AFM ring becomes stronger, leading to enhanced spin asymmetry in transport. Beyond a certain $\gamma$, 
the transport properties become predominantly governed by the ring states, causing the SP to approach saturation.
Similarly, junction 2 exhibits a comparable behavior, as shown in Fig.~\ref{fig:f6}(b). With increasing coupling strength, the magnitude 
of the SP gradually increases. Beyond a certain coupling strength, however, the opposite spin channel starts dominating the transport, 
resulting in a reversal of the spin polarization and the emergence of a large negative SP. This sign reversal indicates a change in the 
dominant spin-transport channel induced by the coupling-dependent hybridization between the NM wire and AFM ring states. In this analysis, 
the electric field strength is fixed at $\varepsilon=0.74$.

To verify the consistency of our results over a wider range of system sizes, we plot the central quantity, namely the spin polarization, 
as a function of the in-plane electric field for different system configurations. The results for junction 1 with $N_R=20$ and $N_W=13$ 
are shown in Fig.~\ref{fig:f7}(a), while those for junction 2 with $N_R=20$ and $N_W=14$ are presented in Fig.~\ref{fig:f7}(b).
For both junction configurations, we fix the bias voltage and Fermi energy at $V=0.2\,$V and $E_F=-0.8\,$eV, respectively. All the key
signatures and essential features remain qualitatively unchanged.

The SP as a color map in the parameter spaces of $\gamma$-$\varepsilon$, $\varepsilon$-$E_F$, and $E_F$-$V$ for junction 1 and junction 2 
are shown in Figs.~\ref{fig:density}(a)-(c) and  Figs.~\ref{fig:density}(d)-(f) respectively. For the $\gamma$-$\varepsilon$ density plot, 
the Fermi energy is fixed at $E_F=-0.5\,$eV for junction 1  and $E_F=-0.8\,$eV for junction 2. The extended red and blue regions, 
representing large positive and negative SP, respectively, confirm that the high degree of spin polarization is not restricted to specific
parameter values but persists over broad ranges of coupling strength, electric field, Fermi energy, and bias voltage. The results for both
junctions consistently show that significant SP can be achieved and controlled through the interplay of ring-wire coupling and the in-plane
electric field, establishing the robustness of the proposed spin-filtering mechanism.
\begin{figure*}[htbp]
	\centering
	\begin{minipage}{0.32\textwidth}
		\centering
		\includegraphics[width=\linewidth,height=4.2cm]{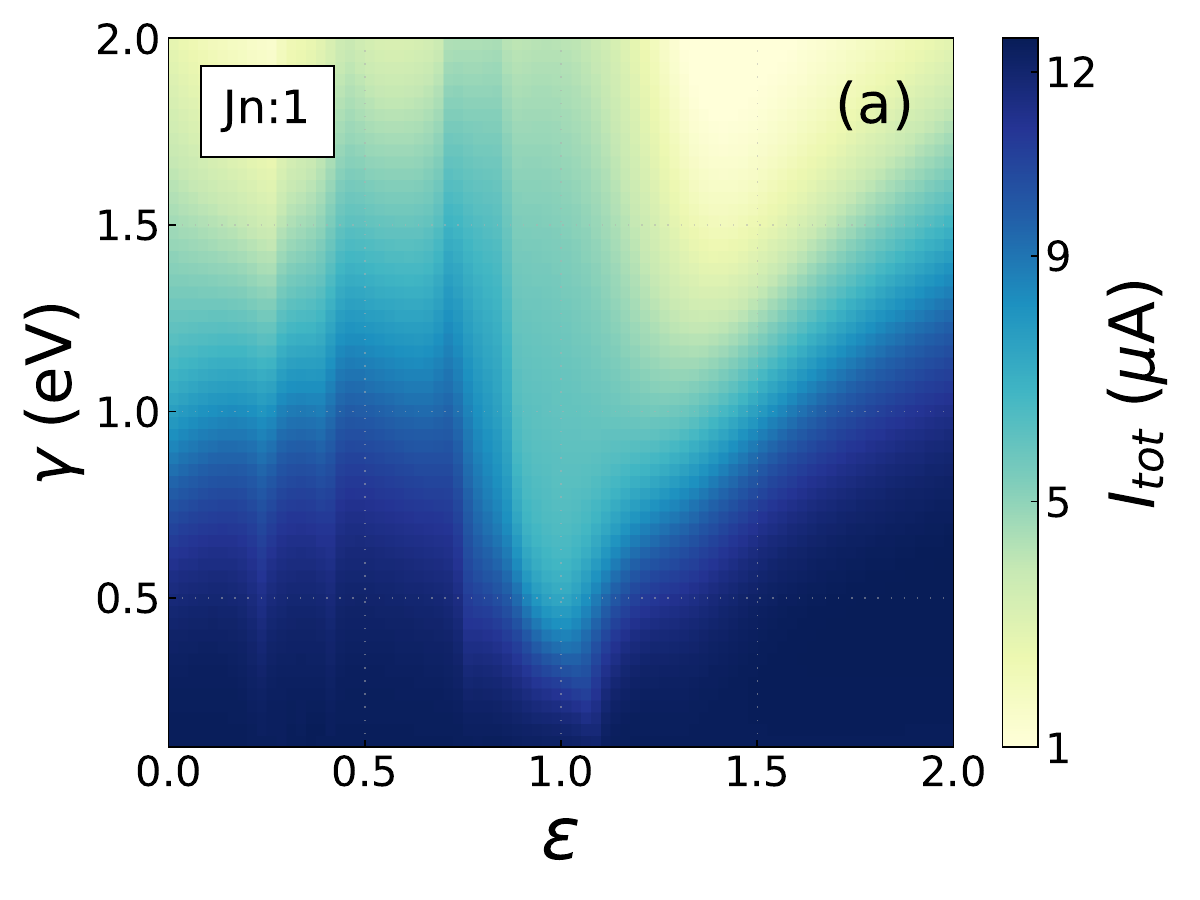}
	\end{minipage}
	\hfill
	\begin{minipage}{0.32\textwidth}
		\centering
		\includegraphics[width=\linewidth,height=4.2cm]{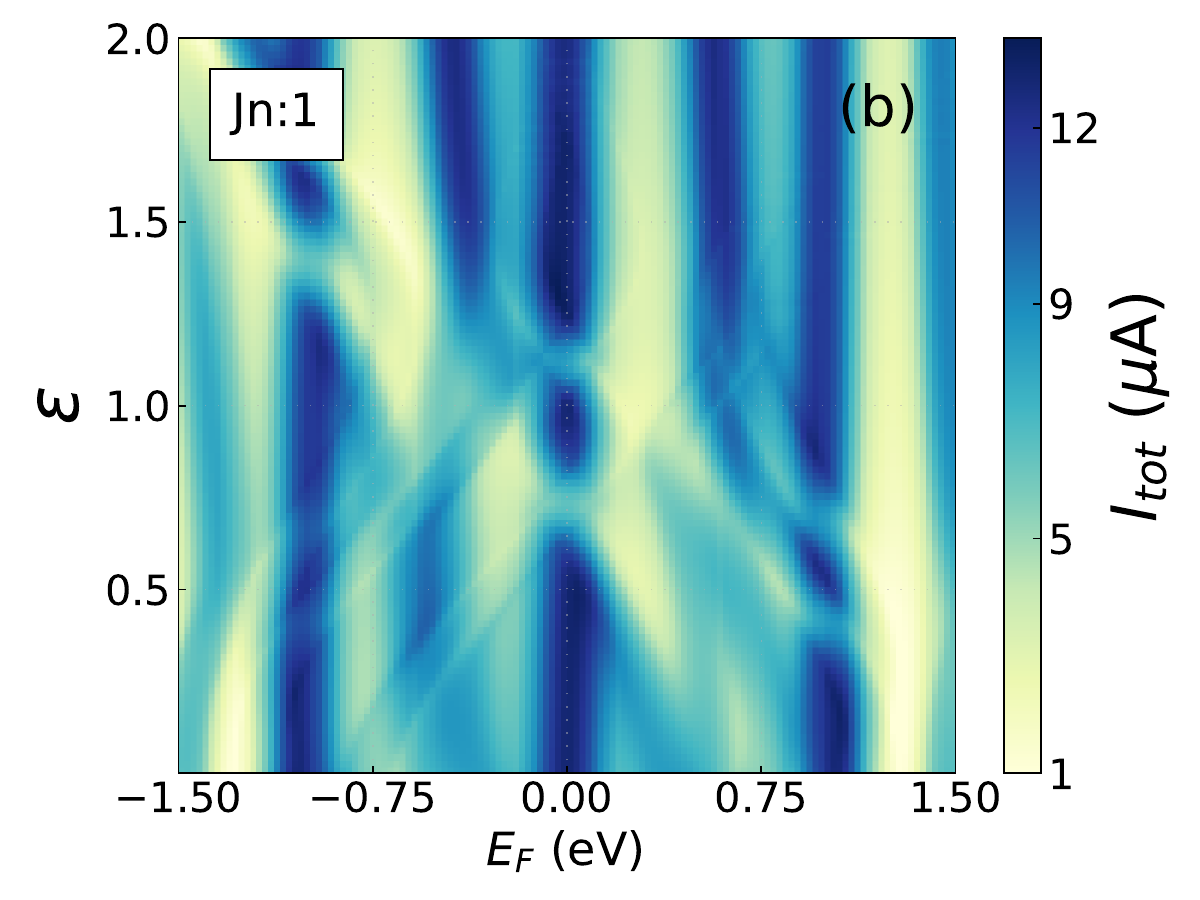}
	\end{minipage}
	\hfill
	\begin{minipage}{0.32\textwidth}
		\centering
		\includegraphics[width=\linewidth,height=4.2cm]{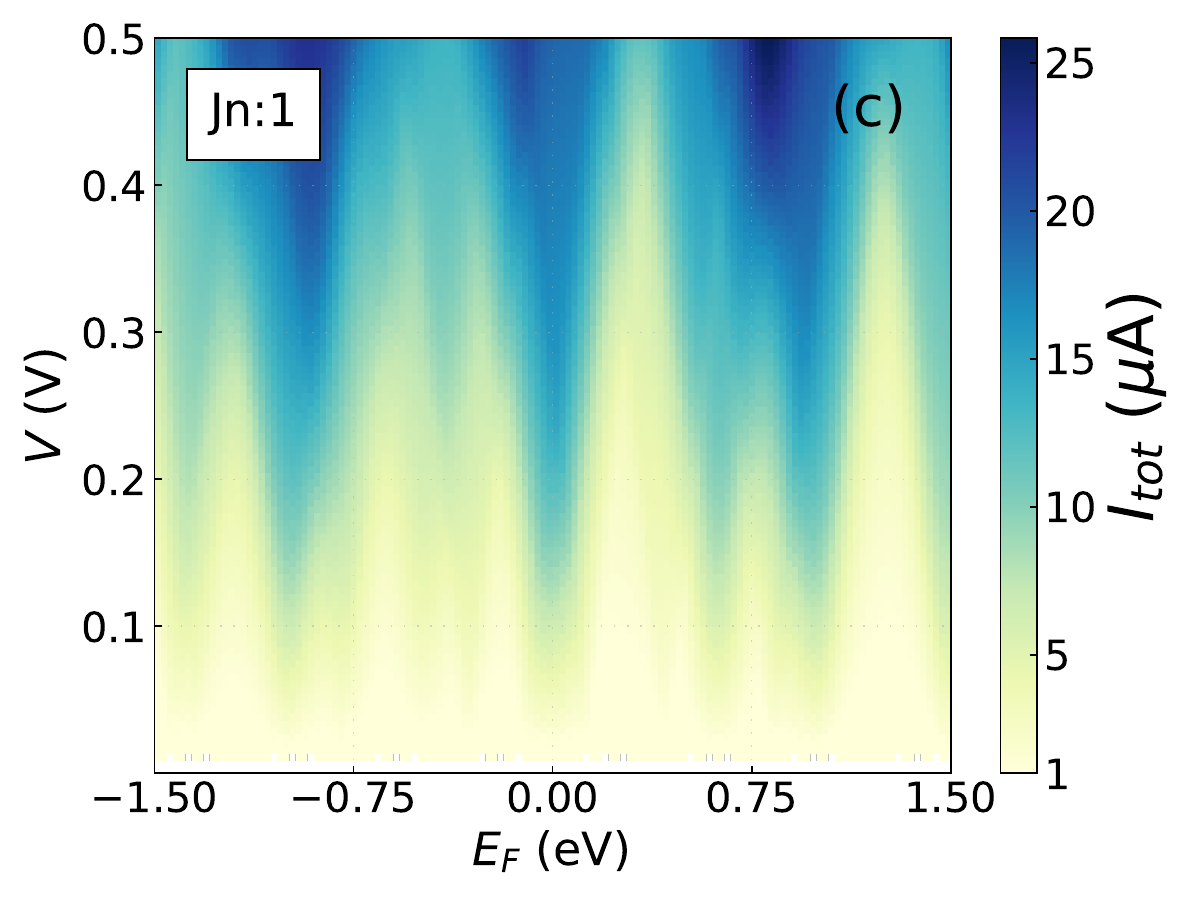}
	\end{minipage}
	
	\vspace{0.2cm}
	
	\begin{minipage}{0.32\textwidth}
		\centering
		\includegraphics[width=\linewidth,height=4.2cm]{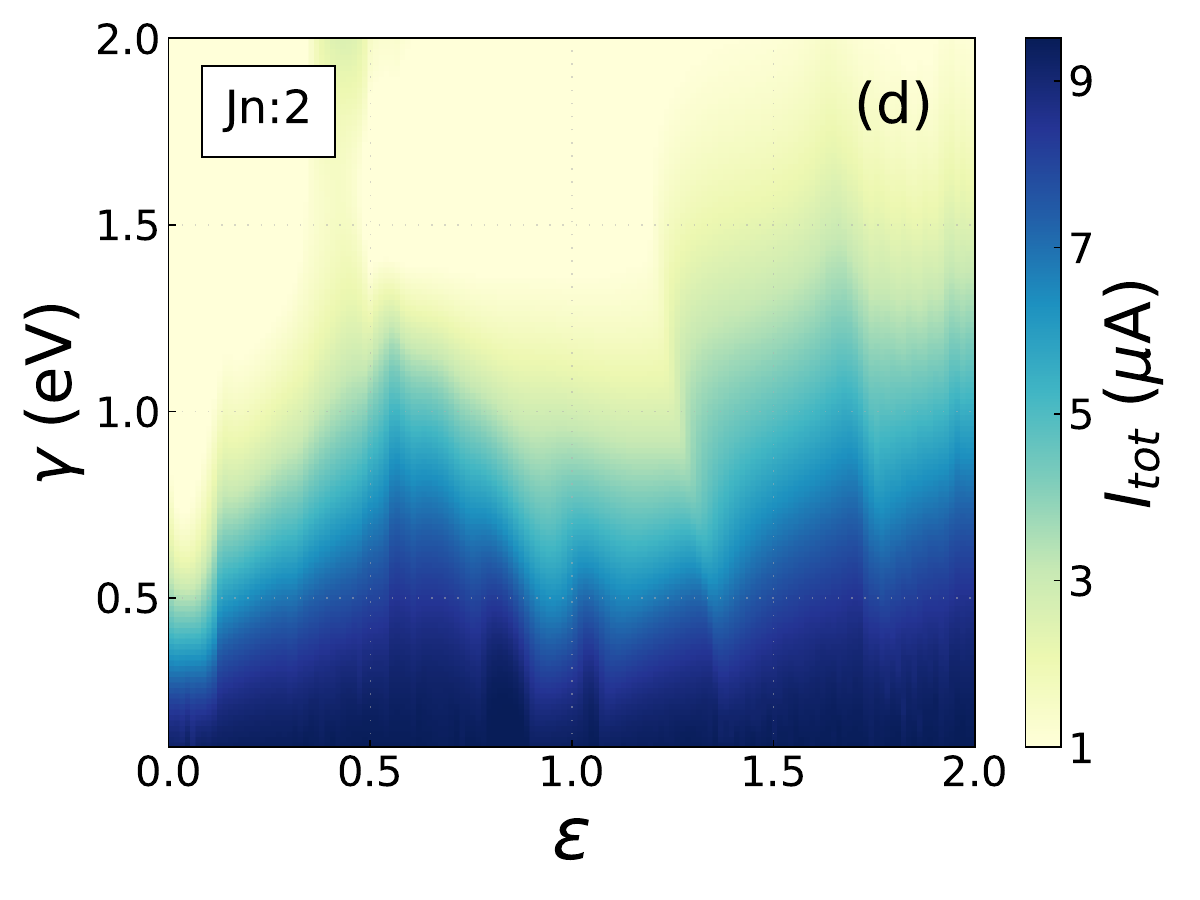}
	\end{minipage}
	\hfill
	\begin{minipage}{0.32\textwidth}
		\centering
		\includegraphics[width=\linewidth,height=4.2cm]{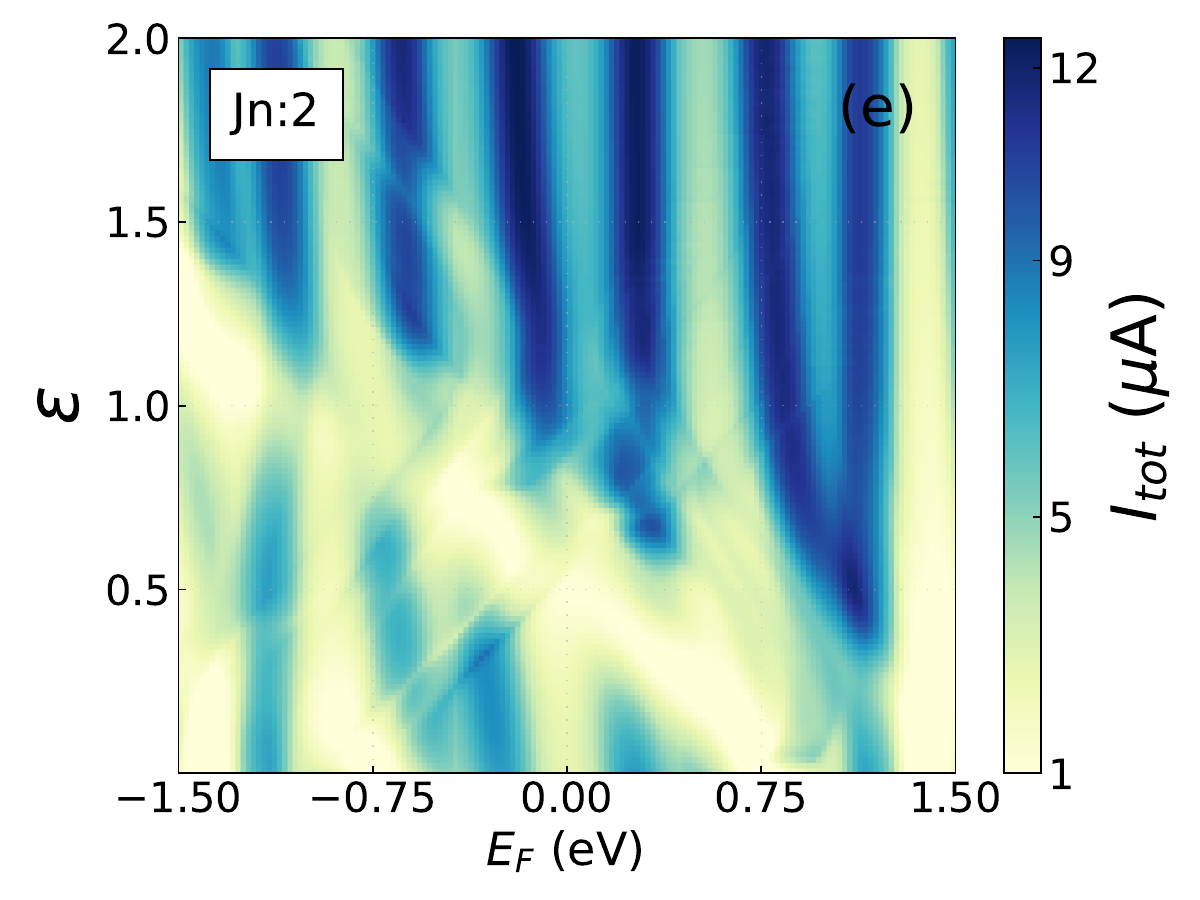}
	\end{minipage}
	\hfill
	\begin{minipage}{0.32\textwidth}
		\centering
		\includegraphics[width=\linewidth,height=4.2cm]{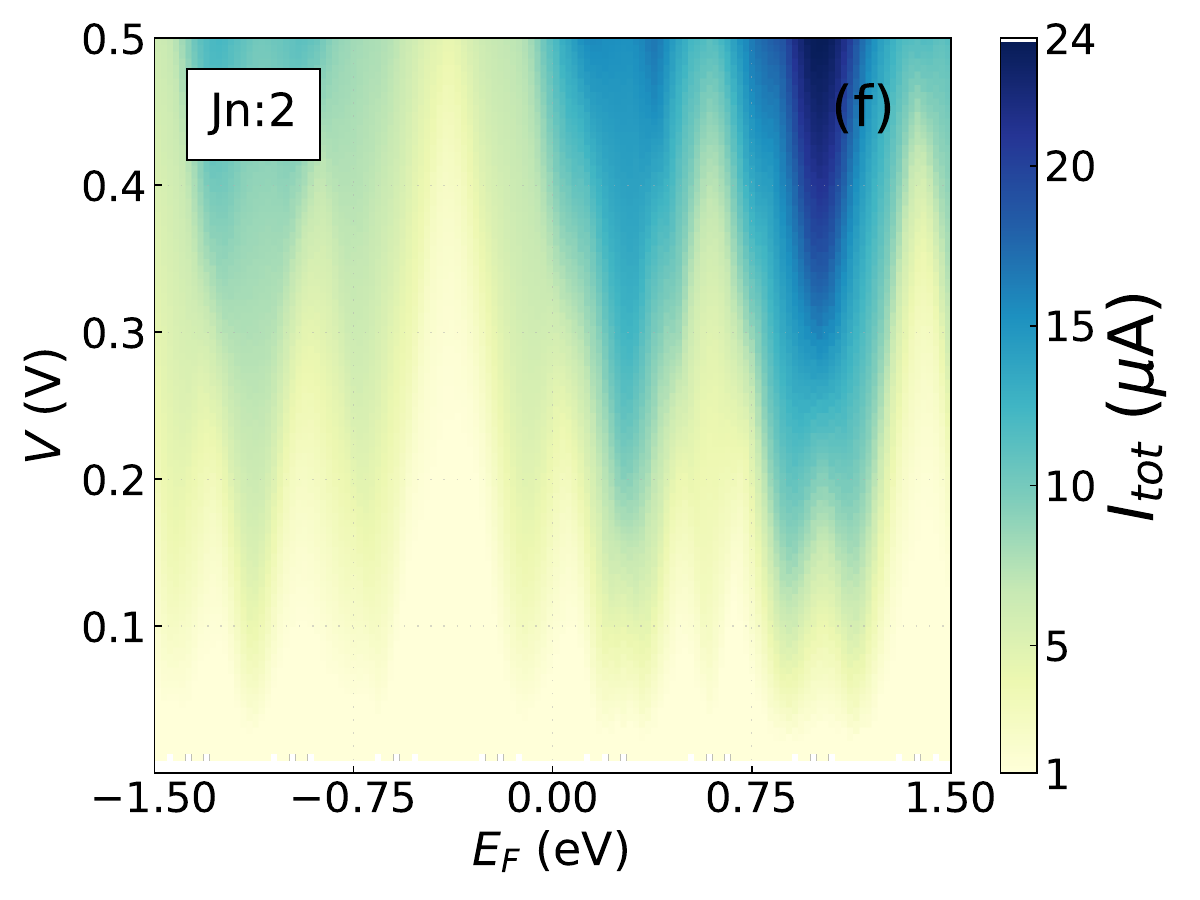}
	\end{minipage}
\caption{(Color online). Density plots of `total' current for different parameter regimes in junction 1 and junction 2. The variation 
of total current with ring-wire coupling and electric field is shown in (a) and (d), with Fermi energy and electric field in (b) and (e), 
and with Fermi energy and bias voltage in (c) and (f). Figures (a)-(c) correspond to junction 1, while figures (d)-(f) represent junction 2.}
\label{fig:cur_density}
\end{figure*}
\begin{figure}[htbp]
	\centering
	\resizebox{7cm}{4.5cm}{\includegraphics{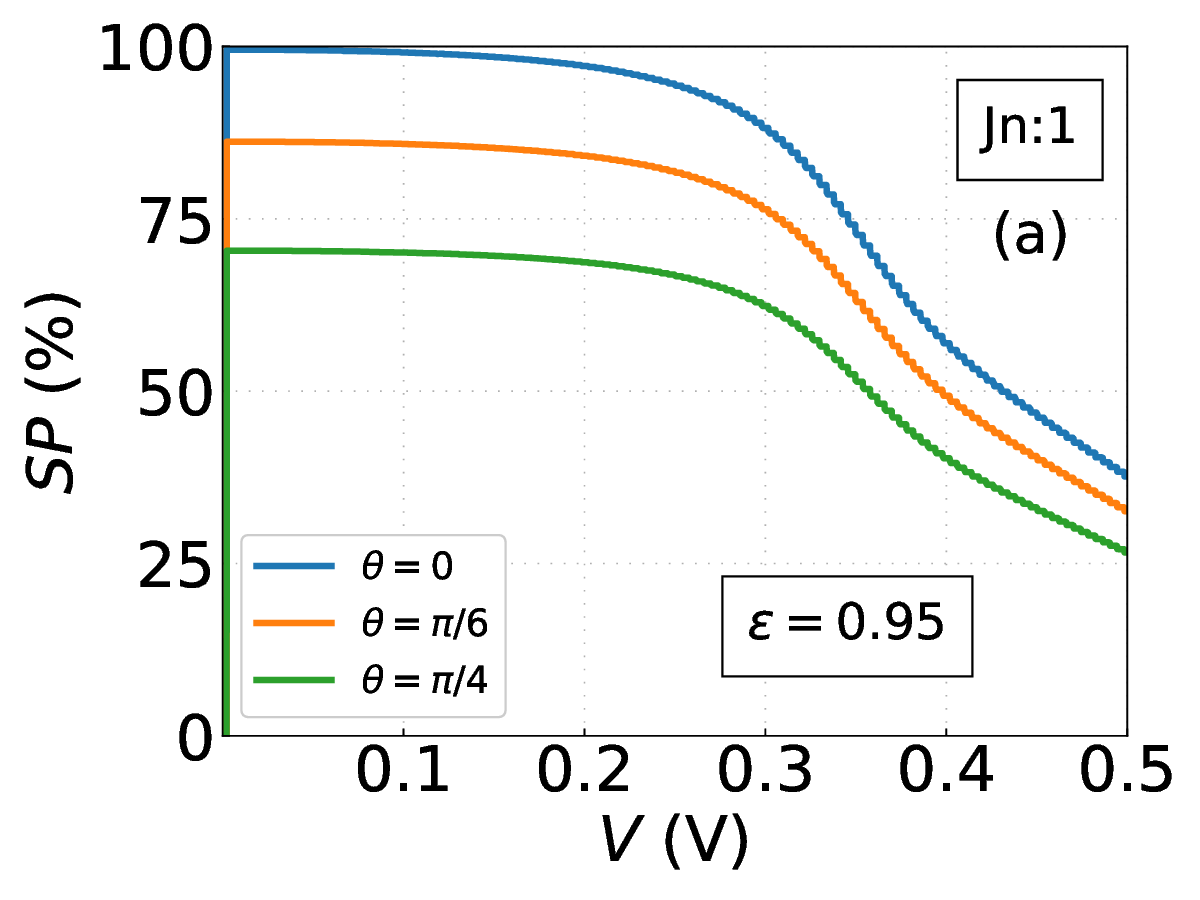}}
	
	\vspace{0.3cm}
	
	\resizebox{7cm}{4.5cm}{\includegraphics{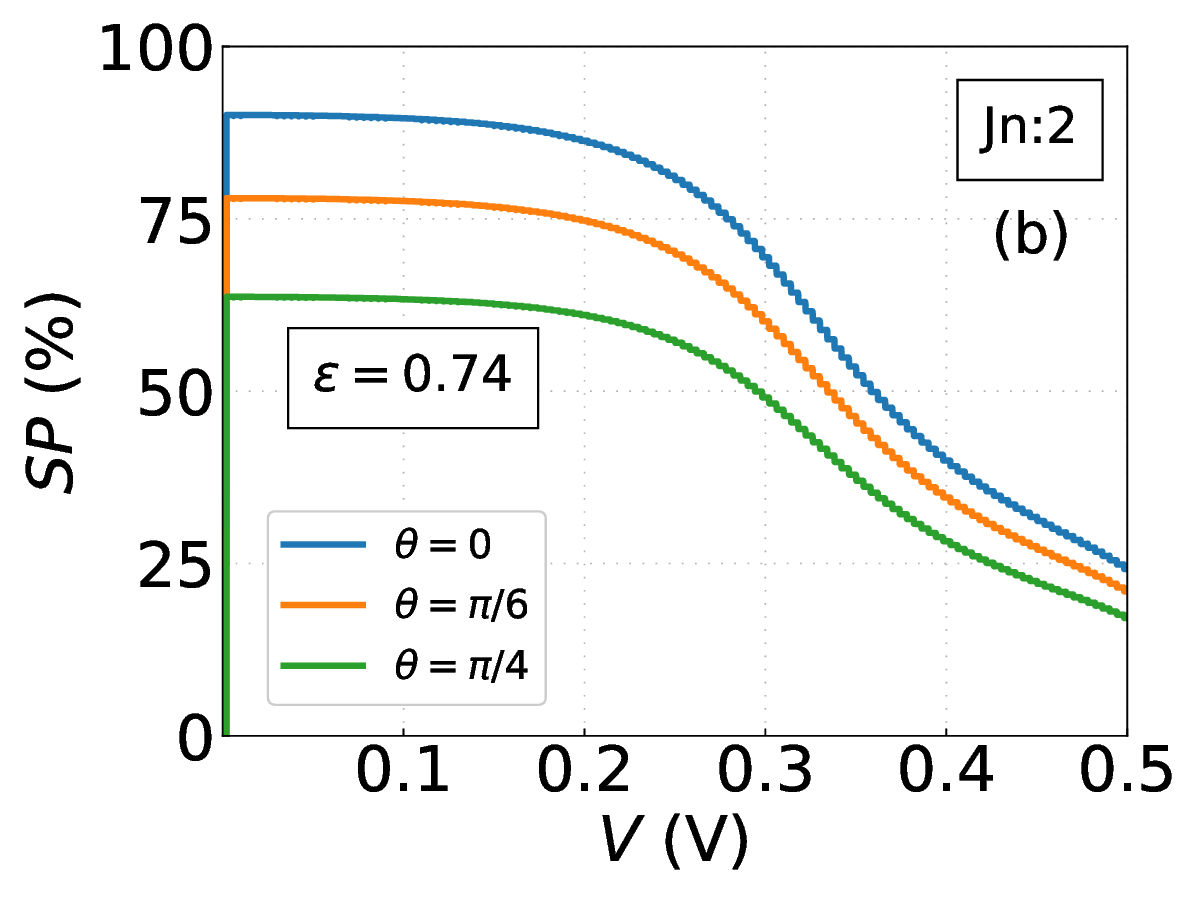}}
	\caption{(Color online). Spin polarization as a function of applied bias voltage for three different canting angles, $\theta=0$, $\pi/6$, 
		and $\pi/4$, for (a) junction 1 at $\varepsilon=0.95$ and (b) junction 2 at $\varepsilon=0.74$.}
	\label{fig:canting}
\end{figure}

To assess whether the obtained SP is significant or merely a numerical artifact, it is important to check the total current 
of the coupled system, since the SP is defined as a ratio of the two spin currents, and values close to $100\%$ can, in principle, occur even
when the total current is extremely small. The total current, shown as color maps in the same parameter spaces as Fig.~\ref{fig:density}, for
junction 1 and junction 2 is presented in Figs.~\ref{fig:cur_density}(a)-(c) and Figs.~\ref{fig:cur_density}(d)-(f), respectively. The plots
clearly show that, over a wide range of parameters, the total current remains appreciable in regions exhibiting large spin polarization 
(see Fig.~\ref{fig:density}).

So far, all the results are obtained by considering perfectly collinear magnetic moments, with the moments oriented 
alternately along the quantized $\pm Z$ directions. With this assumption, the spin-dependent scattering matrix remains diagonal in the 
chosen spin basis, and therefore spin-flip scattering does not occur. We now rotate the magnetic moments with respect to the quantization 
axis by changing $\theta_n$ while keeping the azimuthal angle $\phi_n=0$. With the introduction of the polar angle, the transverse components
$(h_x,h_y)$ of $\boldsymbol{\vec{h}_n}$ become nonzero. These components mix the spin up and spin down states and therefore introduce
spin-flip scattering. The resulting spin mixing is expected to reduce the spin selectivity of the junction. The extent of this effect 
depends on both the magnitude of the canting and its spatial pattern. To inspect this effect, we introduce a set of canting angles
$(\theta=0,\pi/6,\pi/4)$ for all the magnetic moments of the AFM ring and calculate the spin polarization as a function of voltage for 
both junctions, as shown in Figs.~\ref{fig:canting}(a) and (b). We find that the SP is reduced compared with the perfectly collinear case. 
This reduction occurs because the canting introduces off-diagonal terms in the magnetic part of the Hamiltonian, which mix the up and 
down spin states and consequently reduce the spin selectivity of the junction. We also observe that the SP decreases gradually as the 
canting angle is increased, since the strength of this spin mixing becomes larger with increasing canting. It should be noted that the 
azimuthal angle is kept fixed here, since the SP is found to be independent of the azimuthal angle.

In the present work, the in-plane electric field plays a crucial role in spin splitting and its tuning. Although we 
present several results showing the effect of the electric field, its quantitative role in terms of field strength is not discussed. 
In our model, the applied electric field changes the onsite energies of the AFM ring sites, while the hopping parameters are kept fixed.
Therefore, the ring does not suddenly decouple from the wire at any particular field strength. Instead, as the field strength increases, 
the onsite-energy modulation becomes stronger, and the contribution of the AFM ring to the transport is gradually suppressed due to the localization.
\begin{figure}[htbp]
\centering \resizebox{8cm}{5.5cm}{\includegraphics{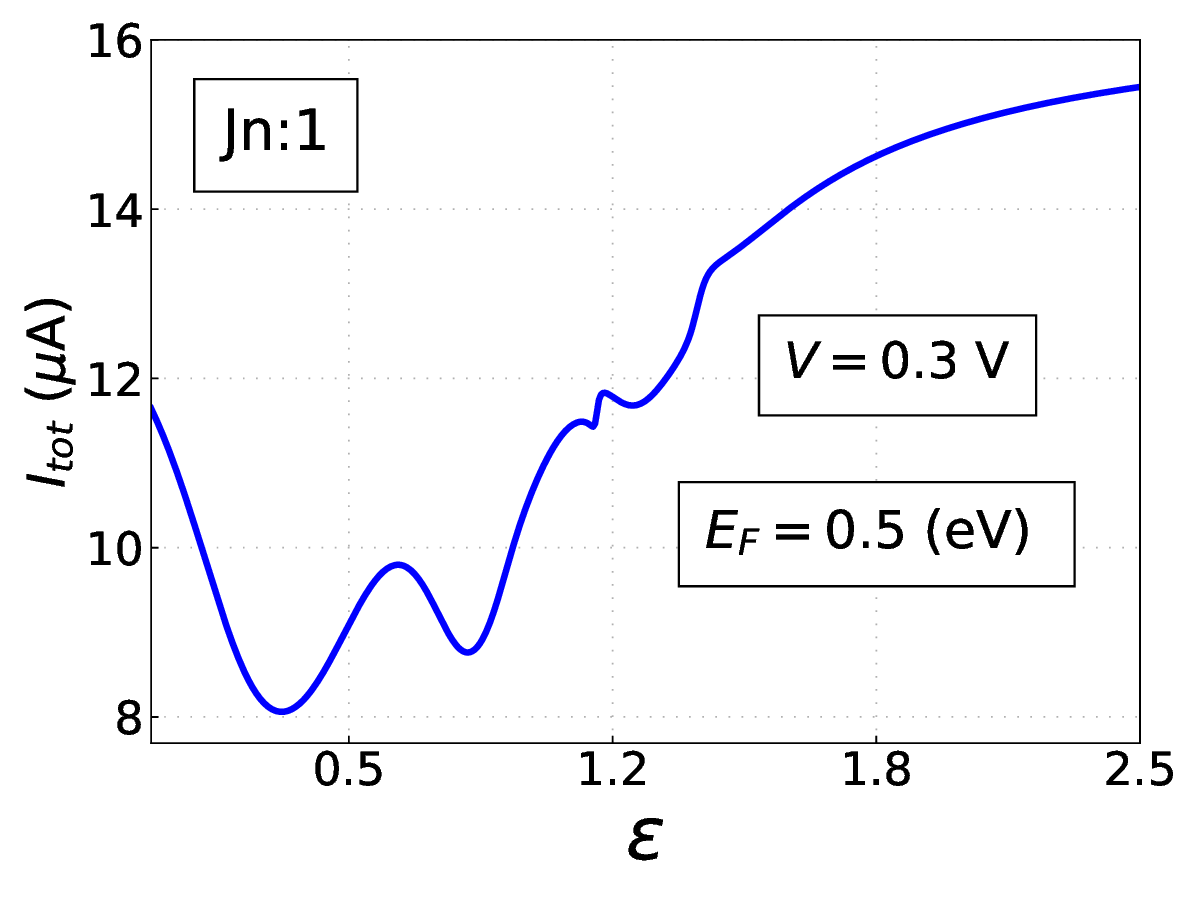}}
\caption{(Color online). Variation of total current with the in-plane electric field at a constant bias voltage $V=0.3\,$V and Fermi 
energy $E_F=0.5\,$eV for junction 1.} 
\label{fig:field}
\end{figure}
\begin{figure}[htbp]
\centering\resizebox{7cm}{5.5cm}{\includegraphics{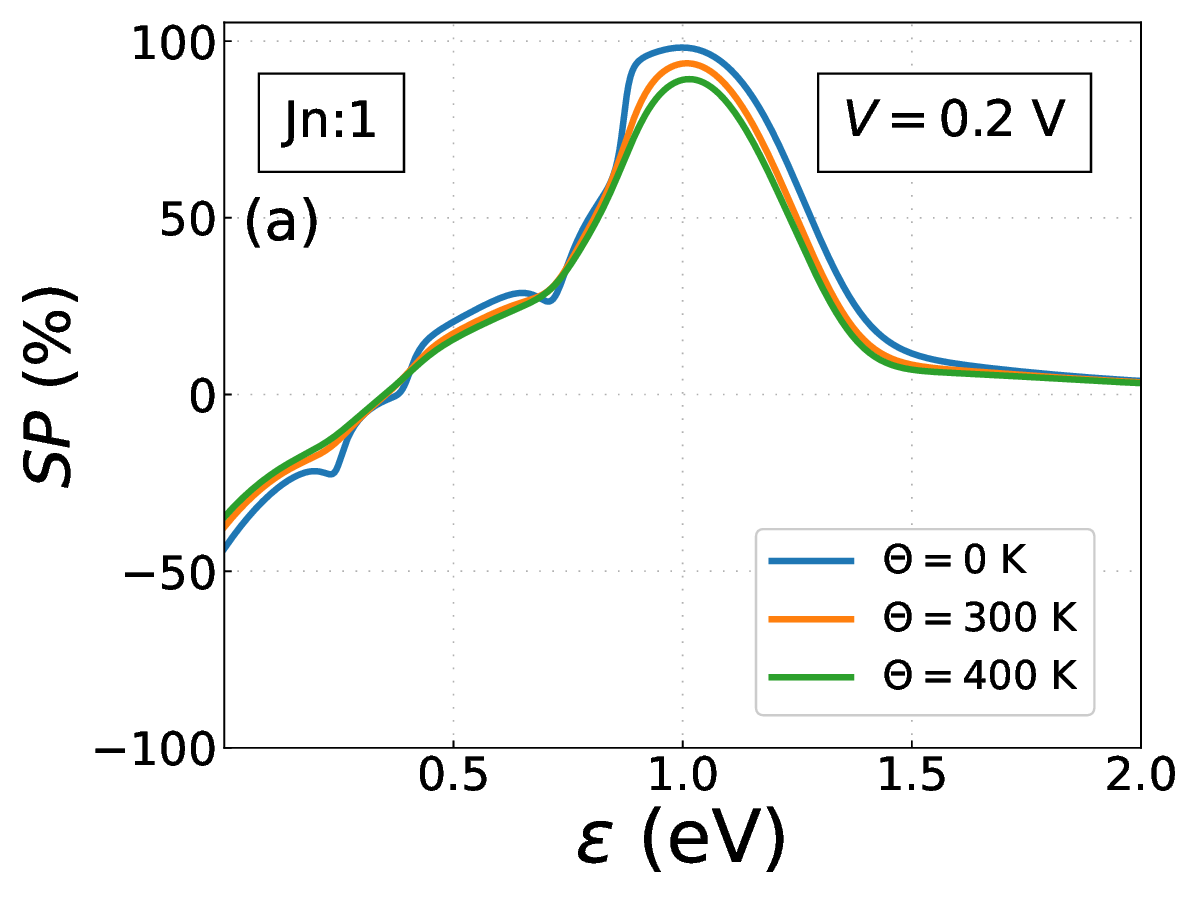}}
	
	\vspace{0.3cm}
	
\resizebox{7cm}{5.5cm}{\includegraphics{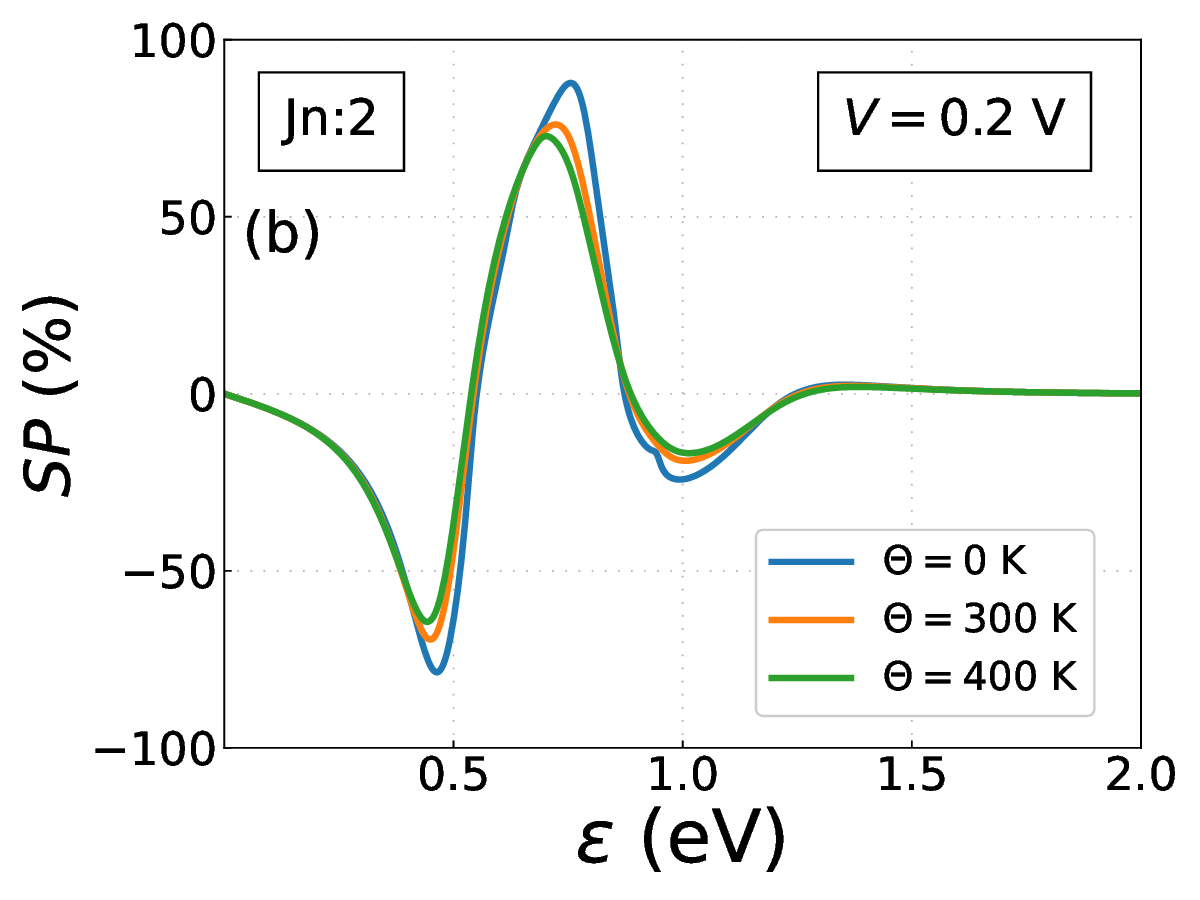}}
\caption{(Color online). Variation of spin polarization with the in-plane electric field for different temperatures $(\Theta=0,\ 300,\ 400\,\text{K})$ at a fixed bias voltage $V=0.2\,$V. Figures (a) and (b) correspond to junction 1 and junction 2, respectively.}
\label{fig:f8}
\end{figure}
To obtain an estimate of the field strength at which the AFM ring starts to decouple from the NM wire, we compute the 
total current as a function of the electric-field strength, while keeping the applied voltage and Fermi energy fixed at $V=0.3\,$V 
and $E_F=0.5,$eV, respectively, as shown in Fig.~\ref{fig:field}. We find that the current initially decreases with increasing field 
strength due to the increasing modification of the ring-site energies. After reaching a minimum, the current starts to increase, indicating 
that the contribution from the NM wire is becoming dominant as the AFM ring gradually decouples from the wire. In the intermediate field 
range, approximately $0.5\lesssim\varepsilon\lesssim0.9$, the current shows weak oscillatory behavior, suggesting that the transport is 
still influenced by the strongly modulated AFM ring while the contribution from the NM wire is gradually becoming more dominant. With a 
further increase in the field strength, the current approaches a saturation value characteristic of transport through the clean 
NM wire. Thus, the field strength corresponding approximately to the minimum in the total current can be regarded as an indication of 
the beginning of ring-wire decoupling, rather than as a sharp or complete decoupling point.

Up to this point, all the presented results correspond to zero temperature ($\Theta=0$). The inclusion of finite temperature does not 
alter the essential characteristics of the observed behavior, as illustrated in Fig.~\ref{fig:f8}. The dependence of the SP on electric 
field is presented for three different temperatures, $\Theta=0$, $300$, and $400\,$K, covering both junction 1 and junction 2 in
Fig.~\ref{fig:f8}(a) and Fig.~\ref{fig:f8}(b), at a constant bias voltage $V=0.2\,$V. Only marginal changes are observed with increasing
temperature, which is itself a favorable outcome. The slight reduction in the magnitude of SP at higher temperatures originates from 
thermal broadening introduced through the difference between the Fermi distribution functions, $(f_S-f_D)$. Since the system size is 
relatively small, the effect of thermal broadening remains very weak and therefore does not significantly modify the overall transport
characteristics.

\section{Experimental perspective}

Finally, we briefly comment on the possible experimental realizations of the proposed system. Recent advances in nanoscale fabrication
techniques have enabled the realization of antiferromagnetic quantum spin rings with atomic precision. Very recently, Li \textit{et. al.}
experimentally\cite{exp1} fabricated antiferromagnetic $S=1/2$ quantum spin rings on an Au(111) surface through on-surface synthesis 
combined with STM-tip-induced manipulation, where neighboring spin sites exhibited strong AFM exchange coupling. Earlier, molecular
antiferromagnetic rings such as Cr-based spin rings\cite{exp2,exp3} were also successfully synthesized and characterized experimentally,
confirming stable antiferromagnetic interactions between neighboring magnetic sites.

The hybrid ring-wire geometry considered in the present work may be realized by coupling such AFM nanorings to semiconductor quantum 
wires or metallic atomic chains using STM-based atom-manipulation techniques. Furthermore, the in-plane electric field required in our 
model can be externally controlled through gate electrodes\cite{exp4}, which is routinely achievable in nanoscale transport experiments.
In dimensionless units, the maximum field strength used in this work is $\varepsilon \approx 2$, which corresponds to 
a physical field of $ \mathcal{E} = t_r\varepsilon/(ea) =  10\,\varepsilon\,$V/nm for a lattice constant $a=1\,$\AA. Such field strengths 
are comparable to intense local fields already realized in experimental studies\cite{exp5}. Therefore, the proposed setup appears to be experimentally feasible within currently available nanotechnology and spintronics platforms.

\section{Closing remarks}

We investigate and analyze spin-dependent transport in a ring-wire hybrid system, where a non-magnetic wire is connected to two electrodes 
and an antiferromagnetic ring is symmetrically coupled to the NM wire in the presence of an in-plane electric field perpendicular to the 
wire. Two different hybrid junction configurations, distinguished by the number of ring-wire couplings, are considered. In the present 
work, the spin symmetry between $H_{rw,\uparrow}$ and $H_{rw,\downarrow}$ is found to be governed by two factors: the number of ring-wire
couplings and the applied in-plane electric field. Within the tight-binding framework, the NEGF formalism is adopted to evaluate the
transmission probability, and the Landauer-B\"{u}ttiker prescription is used to compute the junction currents.

The outcomes of our analysis reveal a high degree of spin polarization over a wide bias window. The restoration or breaking of spin 
symmetry, depending on the number of ring-wire couplings, is also clearly demonstrated. An odd number of ring-wire couplings breaks the 
spin symmetry, whereas an even number of couplings restores it. The results further demonstrate the external tunability of spin-dependent
transport through the in-plane electric field. To assess the consistency of our results, we check a wide range of parameters and find that 
the key features remain qualitatively unchanged under variations in system size, ring-wire coupling strength, field strength, temperature, 
Fermi energy, and voltage.

Recent progress in STM-based nanofabrication enables the construction of AFM rings through on-surface synthesis on Au (111) substrates. 
Since the in-plane electric field can be controlled by a gate electrode and STM-based atom-manipulation techniques allow for the realization 
of the proposed hybrid ring-wire geometry, the present setup provides a possible platform for exploring electrically controlled spin-dependent
transport at the nanoscale.

\end{document}